\documentclass[11pt,a4paper]{article}

\usepackage{jheppub}
\bibstyle{JHEP}
\usepackage[english]{babel}
\pdfoutput=1
\usepackage{physics}
\usepackage{tikz}
\usepackage{amsmath,amssymb,amsfonts}
\usepackage{mathrsfs}
\usepackage{amsthm}
\usepackage{graphicx}
\usepackage{dsfont}
\usepackage{subfigure}
\usepackage{xcolor}
\usepackage{comment}
\usepackage{soul}
\usepackage{booktabs}

\definecolor{darkred}{rgb}{0.8,0.1,0.1}
\hypersetup{
  colorlinks=true,
  urlcolor=blue,
  linkcolor=darkred,
  citecolor=blue,
  linktoc=page,
  linktocpage=true
}

\definecolor{3dcolor}{rgb}{0.96,0.89,0.76}
\definecolor{4dcolor}{rgb}{0.812,0.851,0.914}

\def\RR{{\mathds{R}}}

\def\cC{{\cal C}}

\DeclareMathOperator{\vol}{vol}

\newcommand{\be}{\begin{equation}}
	\newcommand{\ee}{\end{equation}}
\newcommand{\ba}{\begin{array}}
	\newcommand{\ea}{\end{array}}
\newcommand{\bea}{\begin{equation}\begin{aligned}}
		\newcommand{\eea}{\end{aligned}\end{equation}}

\newcommand{\p}{\partial}

\newcommand{\sgn}{\text{sgn}}

\newcommand{\s}[2]{\sum_{#1}^{#2}}
\newcommand{\deri}[2]{\frac{d #1}{d #2}}

\def\RR{{\mathds{R}}}

\def\cC{{\cal C}}

\newcommand{\mo}{\mathcal{O}}
\newcommand{\man}{\mathcal{N}}
\newcommand{\mac}{\mathcal{C}}

\newcommand{\ml}{\mathcal{L}}

\newcommand{\mai}{\mathcal{I}}

\newcommand{\gym}{g_{\mathrm{YM}}}

\newcommand{\ads}{\text{AdS}}

\newcommand{\ns}{N_5}
\newcommand{\nd}{N_{\rm D5}}
\newcommand{\vef}{V_{\rm{eff}}}

\newcommand{\ap}{\alpha}

\title{One-point functions of the Lagrangian in $\mathcal N=4$ SYM with boundaries, defects and interfaces}

\author[]{Dongming He,} \author[]{Christoph F. Uhlemann,} \author[]{Paolo Vallarino}

\affiliation[]{Theoretische Natuurkunde, Vrije Universiteit Brussel and The International Solvay Institutes, Pleinlaan 2, B-1050 Brussels, Belgium}

\emailAdd{dongming.he@vub.be}
\emailAdd{christoph.uhlemann@vub.be}
\emailAdd{paolo.vallarino@vub.be}

\abstract{
We obtain one-point functions of the ambient Lagrangian $\mathcal O_\ml$ in $\man=4$ SYM with $\frac{1}{2}$-BPS boundaries, defects and interfaces classified by Gaiotto and Witten, which include maximally supersymmetric Janus interfaces and BCFTs relevant for double holography, as well as for Janus interfaces with reduced supersymmetry. 
For Janus interfaces we extend recent results capturing $\langle \mathcal O_{\mathcal L}\rangle$ in certain limits to exact dependence on $N$ and $\lambda$, and confirm a non-renormalization conjecture. We also clarify recent results for Janus variants preserving less supersymmetry.
For general $\frac{1}{2}$-BPS Gaiotto-Witten theories with genuine defect degrees of freedom we derive one-point functions using supersymmetric localization, confirm that the results agree with holographic predictions in the appropriate limits, and show that the one-point functions in general exhibit non-trivial coupling dependence. 
}

\begin{document}
\maketitle
\flushbottom

\section{Introduction and Summary}\label{sec: intro}

The simplest interacting quantum field theories with boundaries, defects or interfaces in 4~dimensions are arguably those with maximal (super)symmetry, classified by Gaiotto and Witten \cite{Gaiotto:2008sd,Gaiotto:2008sa,Gaiotto:2008ak}. They combine 4d $\mathcal N=4$ SYM theories on half spaces, with independent couplings and gauge groups, with junction conditions involving 3d conformal matter as well as Dirichlet, Neumann and Nahm pole boundary conditions. These theories preserve an $SO(2,3)$ subgroup of the 4d conformal symmetry and half the supersymmetries. They enrich 4d $\mathcal N=4$ SYM by interfaces, i.e.\ codimension-1 surfaces separating different 4d theories, defects carrying additional 3d degrees of freedom, and boundaries which terminate the 4d theory. The constructions in particular allow for combinations, e.g.\ interfaces which carry additional 3d matter and (partly) constrain the ambient theories by boundary conditions. We refer to this broad class of theories as 4d Gaiotto-Witten theories.
They are a natural benchmark for our understanding of boundaries, defects and interfaces.

The simplest correlation functions in these theories are one-point functions, and the simplest operators are chiral primaries. They have been studied extensively for theories which are small deformations of standard 4d $\mathcal N=4$ SYM, e.g.\ \cite{Nagasaki:2011ue,Nagasaki:2012re,deLeeuw:2015hxa,Buhl-Mortensen:2015gfd,Buhl-Mortensen:2016pxs,deLeeuw:2017cop,deLeeuw:2019usb,DeLeeuw:2018cal,Linardopoulos:2025ypq,Robinson:2017sup,Wang:2020seq,Komatsu:2020sup}, and more recently also for genuine backreacted deformations \cite{He:2025due}. We will focus on the latter class of fully backreacted deformations. This in particular includes BCFTs, which are never small deformations, and Janus interfaces with sizable coupling differences.

A distinguished operator is the exactly marginal one arising from the Lagrangian. Its one-point functions have been studied using holography and conformal perturbation theory for non-BPS Janus interfaces in \cite{Clark:2004sb} and, more recently, for supersymmetric Janus interfaces in \cite{Karch:2026ymg}. While the Lagrangian seen as a local operator is in principle ambiguous due to integration by parts identities, supersymmetry provides a preferred choice for $\mathcal N=4$ SYM: the fourth descendant in the stress-tensor multiplet. This operator is related to the form of the $\man=4$ SYM Lagrangian with canonical kinetic terms by a total derivative \cite{Chicherin:2016fbj,Bason:2023bin}. This particular version of the Lagrangian operator, $\mathcal O_{\mathcal L}$, is the focus of this work.

For the simplest of the Gaiotto-Witten theories, which is the $\mathcal N=4$ Janus interface with no 3d degrees of freedom separating 4d $\mathcal N=4$ SYM theories with different gauge couplings, a curious property of $\mathcal O_{\mathcal L}$ was noted in \cite{Karch:2026ymg}: the one-point functions computed at leading order in conformal perturbation theory near standard 4d $\mathcal N=4$ SYM agree with holographic results valid in the regime of strong coupling and large $N$, with both depending linearly on the coupling difference. This suggests that the result may in fact be exact. For Janus interfaces with less supersymmetry, the one-point functions obtained holographically from the solutions in \cite{Bobev:2020fon} were found to have complicated coupling dependence.

Here we extend these calculations in two directions, using a combination of supersymmetric localization and holography. Firstly, we provide results for more general Gaiotto-Witten theories, e.g.\ involving interfaces between ambient theories with different gauge groups, various forms of 3d defect degrees of freedom, and boundary CFTs. Our results, which include exact expressions for simple BCFTs and large-$N$ results for more involved boundary, defect and interface CFTs, show that the linear coupling dependence is special to the Janus interface and not a consequence of symmetry alone -- $\langle \mathcal O_{\mathcal L}\rangle$ in general has non-trivial coupling dependence. Secondly, we extend and update the results for Janus interfaces in $\mathcal N=4$ SYM preserving varying degrees of supersymmetry, which differ in their interface terms and were classified in \cite{DHoker:2006qeo}. We obtain exact results for the one-point function in $\man=4$ and $\man=2$ Janus interfaces using supersymmetric localization, and find both to be linear in the coupling difference. For $\mathcal N=4$ Janus this establishes the exactness property suggested in \cite{Karch:2026ymg}. We clarify how the exact $\mathcal N=2$ Janus result is consistent with holography, and update the results for $\mathcal N=1,0$ Janus interfaces.

\begin{center}
	\bf -- Summary of results --
\end{center}

Before transitioning to the main part of the paper we provide a more detailed summary of our results.
For a local scalar operator $\mathcal O$ inserted away from the defect in a B/d/ICFT on flat space, conformal symmetry fixes the spacetime dependence to \cite{McAvity:1995zd,Billo:2016cpy}
\bea \label{eq: one-point-gen}
\expval{\mo(x^{i},x_3)}=\frac{\cC_{\mo}}{x_3^{\Delta_{\mo}}}~,
\eea
where $x^i$ denote the coordinates parallel to the defect, $x_3$ is the transverse distance from it, and $\Delta_{\mo}$ is the scaling dimension of $\mo$.
Once the normalization of the local operator is fixed, the coefficient $\cC_{\mo}$ encodes dynamical information about the underlying theory and for example provides important input for the boundary bootstrap program \cite{Liendo:2012hy,Liendo:2016ymz}. Our results are summarized in Table~\ref{tab:results} and will be discussed in the remainder of this section.

We first compute $\langle\mathcal O_{\mathcal L}\rangle$ in $\man=4$ SYM with various $\frac{1}{2}$-BPS boundaries, defects or interfaces which in particular can host 3d $\man=4$ matter and can backreact. We illustrate the general formalism for the following (classes of) theories:
\begin{itemize}
	\item[--] the $\mathcal N=4$ Janus interface with no defect matter, which separates 4d $\mathcal N=4$ SYM theories with different couplings,
	\item[--] BCFTs realized by D3-branes ending on NS5-branes, which implement boundary conditions where 4d $\mathcal N=4$ SYM couples to a 3d CFT known as $T[SU(N)]$, 
	\item[--] D3/D5 defect CFTs comprising $\mathcal N=4$ SYM coupled to a (possibly large) number of 3d defect hypermultiplets localized to a defect,
	\item[--] BCFTs realized by D3-branes ending on NS5 and D5 branes, where the number of defect degrees of freedom is independent of the ambient central charge. In these theories the defect sector can itself admit a gravitational description, leading to the notion of double holography \cite{Karch:2000ct,Karch:2000gx,Uhlemann:2021nhu,Karch:2022rvr} which has found applications in black hole studies. 
\end{itemize}
The computations are performed using supersymmetric localization \cite{Witten:1988ze,Nekrasov:2002qd,Pestun:2007rz}, where the relevant matrix models can be constructed using gluing formulas \cite{Dedushenko:2018tgx} from the results for the involved 4d and 3d matter, and using holography. The holographic duals are based on the string theory realization of the theories as D3-branes ending on or intersecting combinations of D5 and NS5 branes. This leads to  Type IIB supergravity solutions with warped $\ads_4 \times S^2 \times S^2 \times \Sigma$ geometry which include regions asymptotic to $\ads_5 \times S^5$~\cite{DHoker:2007zhm,DHoker:2007hhe,Aharony:2011yc,Assel:2011xz}. 

\begin{table}[t]
	\centering
	\begin{tabular}{c|c|c}
		\toprule
		Theory & $\mathcal C_{\mathcal O_{\mathcal L}}$ in $\langle \mathcal O_{\mathcal L}\rangle=\mathcal C_{\mathcal O_{\mathcal L}} x_3^{-4}$ & Validity \\
		\midrule
		\rule{0pt}{1.2em} & & \\[-1.1em]
		
		$\mathcal{N}=4$ Janus
		&
		$\displaystyle -\dfrac{3N^2}{16\pi^2}\frac{\lambda_L-\lambda_R}{\lambda_L+\lambda_R}\,\sgn(x^3)$
		&
		exact $N$, $\lambda_{L/R}$
		\\[1.4em]
		
		D3/NS5 BCFT
		&
		$\displaystyle\dfrac{3N^2}{16\pi^2}
		+\dfrac{\lambda}{64 \pi^2 }\left(\frac{N^2}{N_5^2}-1\right)$
		&
		exact $N$, $\lambda$
		\\[1.4em]
		
		small D5 defect
		&
		$\displaystyle-\frac{3\,N\sqrt{\lambda}}{8\,\pi^3}\int_{0}^{\frac{\pi}{2}} \!d\theta\,\sin^2\!\theta \cos\theta\tanh\!\left(\frac{\sqrt{\lambda}}{2}\cos\theta  \right) $
		&
		$N\gg 1$, exact $\lambda$ 
		\\[1.4em]
		
		large D5 defect & $\dfrac{N^2}{16\pi^2}(d^2-1)(d^2+3)$ & $N,\lambda\gg 1$
		\\[1.4em]
		
		D3/D5/NS5 BCFT & $\displaystyle -\frac{N^2}{64\pi^2}(\mathfrak{N}_{\mathrm{D}5} - \mathfrak{N}_5)
		\frac{12(\mathfrak{N}_{\mathrm{D}5} + \mathfrak{N}_5)^2+1}{(\mathfrak{N}_{\mathrm{D}5} + \mathfrak{N}_5)^3}$ & $N,\lambda\gg 1$
		\\[1em]
		
		\midrule
		\rule{0pt}{1.2em} & & \\[-1.1em]
		
		$\mathcal{N}=2$ Janus & $\displaystyle -\dfrac{3N^2}{16\pi^2}\frac{\lambda_L-\lambda_R}{\lambda_L+\lambda_R}\,\sgn(x^3)$ & exact $N$, $\lambda_{L/R}$
		\\[1em]
		$\mathcal N=1$ Janus& summarized in (\ref{eq:N1Janus-cO-sum}) & $N,\lambda_{L/R}\gg 1$
		\\[1em]
		$\mathcal N=0$ Janus & $\displaystyle\gamma=\tanh\left[\frac{2c}{3} \,
		_4F_3\left(\frac{1}{2},\frac{3}{4},1,\frac{5}{4};\frac{5}{6},\frac{7}{6},\frac{3}{2};\frac{32}{81}c^2\right)\right]$ & $N,\lambda_{L/R}\gg 1$
		\\[1em]
		\bottomrule
	\end{tabular}
	\caption{One-point functions $\langle \mathcal O_{\mathcal L}\rangle$ for a sample of Gaiotto-Witten theories and for $\mathcal{N}=2,1,0$ Janus interfaces, with their range of validity. For the large D3/D5 defect and the D3/D5/NS5 BCFT,
		$d=\sqrt{1+\mathfrak{N}_{\mathrm{D5}}^2}-\mathfrak{N}_{\mathrm{D5}}$ and $\mathfrak{N}_{\mathrm{D5}}=N_{\mathrm{D5}}\sqrt{\lambda}/(4\pi N)$, $\mathfrak{N}_5 = N_5/\sqrt{\lambda}$. The results are exact in $\mathfrak{N}_5$ and $\mathfrak{N}_{\rm D5}$. For $\mathcal N=0$ Janus, the one-point function $c=-\sgn(x^3)\,8\pi^2N^{-2}\mathcal C_{\mathcal O_{\mathcal L}}$ is fixed by the given relation in terms of $\gamma=(\lambda_L-\lambda_R)/(\lambda_L+\lambda_R)$. 
	} \label{tab:results}
\end{table}

Our field theory computations are based on connecting the one-point function of the Lagrangian, $\langle\mathcal O_{\mathcal L}\rangle$, to the one-point function of the lowest chiral primary operator $\langle\mathcal O_2\rangle$ and to a derivative of the defect free energy. 
Since the 4d $\mathcal N=4$ superconformal symmetry where $\mathcal O_2$ and $\mathcal O_{\mathcal L}$ share a multiplet is broken in the theories of interest here, a relation is not immediate.
Building on the results of \cite{Goto:2020per}, however, we show that $\langle \mo_{\mathcal L}\rangle$ and $\langle \mathcal O_2\rangle$ are still related in theories which preserve 3d $\mathcal N=2$ supersymmetry and where the complex coupling of the ambient theory is in addition allowed to be (smoothly) position-dependent.
Using supersymmetric localization, $\langle\mathcal O_2\rangle$ can be obtained as local one-point function on the (hemi-)sphere, which can in turn be related to the one-point function on flat space. 
This allows us to obtain $\expval{\mathcal O_{\mathcal L}}$ from $\expval{\mo_2}$ for Gaiotto-Witten theories and for $\mathcal N=2$ and $\mathcal N=4$ Janus interfaces, where allowing for smoothly position-dependent coupling becomes relevant. 
Computing $\expval{\mo_2}$, in turn, is operationally equivalent to computing the derivative of the defect free energy on the (hemi-)sphere with respect to the Yang-Mills or 't Hooft coupling, so that we can also obtain $\langle\mathcal O_{\mathcal L}\rangle$ from the defect free energy.

The defect free energies can be obtained non-perturbatively for the empty Janus interface, and have been computed non-perturbatively for the D3/NS5 BCFTs in \cite{Raamsdonk:2020tin}. This allows us to obtain $\langle\mathcal O_{\mathcal L}\rangle$ non-perturbatively. For more general theories engineered by combinations of D5 and NS5 branes in a holographic limit with backreacted 5-branes, the saddle points dominating the matrix models were derived in \cite{He:2024djr} and the one-point functions of chiral primary operators $\mo_J$ were computed in \cite{He:2025due}, again allowing us to obtain $\langle\mathcal O_{\mathcal L}\rangle$.

We complement the localization results with holographic computations, deriving the one-point functions directly on flat space from the asymptotic expansion of the dilaton in the aforementioned holographic duals. This provides consistency checks for our localization results and non-trivial tests of the dualities. The dualities with fully backreacted 5-branes have been validated previously through comparisons of non-local quantities like defect free energies and Wilson loops, by matching aspects of the spectrum, and the supergravity solutions have been connected directly to the matrix models arising in supersymmetric localization \cite{Assel:2012cp,Bachas:2017wva,Raamsdonk:2020tin,Coccia:2020wtk,Chaney:2024bgx,He:2024djr}. The computations here extend these checks to local operators.

The main results are summarized in the first part of Table \ref{tab:results}. The exact result for the $\mathcal N=4$ Janus interface extends the results obtained in \cite{Karch:2026ymg} and confirms the exactness conjecture. We give a scaling argument deriving this directly from the matrix model in section \ref{sec:non-ren}, which also highlights how non-trivial one-loop determinants, as generically present in theories with 3d matter, spoil the scaling behavior and lead to non-trivial coupling dependence. This is manifest in the results for the more general theories involving genuine 3d degrees of freedom in Table \ref{tab:results}, where $\expval{\mathcal O_{\mathcal L}}$ generally has non-trivial coupling dependence.

In a second part we discuss Janus interfaces with reduced supersymmetry. As we will discuss, the localization discussion extends to $\mathcal N=2$ interfaces, and the matrix models capturing supersymmetric observables are identical for the $\mathcal N=4$ Janus interface and the $\mathcal N=2$ Janus interface with $SU(2)$ flavor symmetry identified in \cite{DHoker:2006qeo}. This implies that $\langle \mathcal O_{\mathcal L}\rangle$ is identical in its exact, simple dependence on $N$ and $\lambda$, as reported in Table~\ref{tab:results}.
We then turn to holographic descriptions of interfaces with $\mathcal N=2$ and $\mathcal N=1$ supersymmetry, constructed in \cite{Bobev:2020fon}, and compute the one-point functions.\footnote{This was motivated by the holographic computations in the initial version of \cite{Karch:2026ymg}, which found non-trivial coupling dependence for the $\mathcal N=2$ interface with $SU(2)$ flavor symmetry. Following discussions with the authors, the results in \cite{Karch:2026ymg} have subsequently been revised and will appear together with this preprint.}
We uplift the 4d supergravity solutions of \cite{Bobev:2020fon} to 10d Type IIB supergravity using the uplift formulae provided there, and then find results in agreement with localization for the $\mathcal N=2$ interface. Extending the localization arguments to $\mathcal N=1$ interfaces is less straightforward, but we provide a holographic result for the $\mathcal N=1$ Janus interface with $SU(3)$ flavor symmetry. 
In this case we find non-trivial coupling dependence. Finally, we include the one-point function for the non-supersymmetric Janus interface of \cite{Bak:2003jk,Clark:2004sb}. For all aforementioned interfaces we also compute the defect free energy, and find the same relation to $\langle \mathcal O_{\mathcal L}\rangle$ as in the maximally supersymmetric case. As we discuss in the main part, this can be seen as a consistency check for the non-supersymmetric Janus Lagrangian proposed in~\cite{Clark:2004sb}.

\textbf{Outline}:
In section \ref{sec: local} we review the Lagrangian operator and relate its one-point function to the derivative of the defect free energy with respect to the ambient coupling. We also relate it to the one-point function of the lowest chiral primary operator for theories preserving 3d $\mathcal N=2$ supersymmetry. In section \ref{sec: SYM} we apply these relations to extract the Lagrangian one-point function for 4d Gaiotto-Witten theories using supersymmetric localization. The sample of theories includes the empty $\mathcal{N}=4$ Janus interface and the D3/NS5 BCFT, for which we obtain exact results, as well as the D3/D5 defect and theories with more general (unquenched) 3d degrees of freedom, for which we obtain large-$N$ results. We also discuss non-renormalization properties in that section. In section \ref{sec: holo} we compute the one-point functions for Gaiotto-Witten theories holographically, leading to a general expression, and compare to the localization results. In section \ref{sec:Janus-less-susy} we discuss Janus interfaces with $\mathcal N<4$ supersymmetry, using a combination of localization arguments and holography. This way we obtain the one-point functions for interfaces with $\mathcal N=2,1,0$ supersymmetry.

\section{One-point function of the Lagrangian \texorpdfstring{$\langle \mathcal O_{\mathcal L}\rangle$}{}}\label{sec: local}

In this section we first review relevant aspects of the operator we are computing the one-point functions for. We then discuss relations of its one-point function to chiral primary operators and to the defect free energy, which will facilitate the computations.

Our target is the one-point function of the exactly marginal operator arising from the Lagrangian of 4d $\mathcal N=4$ SYM interface, boundary and defect CFTs. More precisely, this refers to the Lagrangian associated with the ambient theory, away from the defect and without contributions from possible defect degrees of freedom. In general, the local form of the Lagrangian is not unique, and changes e.g.\ under integration by parts transformations in the action, which trade a change in the ambient Lagrangian for boundary terms. Different choices, and the resulting one-point functions, were recently explored in \cite{Karch:2026ymg}.

For supersymmetric theories there are preferred choices. In 4d $\mathcal N=4$ SYM, the natural choice is the operator in the stress-energy tensor multiplet, whose bottom component is the conformal-dimension-two chiral primary operator $\mathcal O_2$. 
General chiral primary operators of conformal dimension $J$ can be defined as
\begin{align}\label{eq:OJ-gen}
\mathcal{O}_J &\sim \mathrm{tr} \left(  u \cdot \Phi \right)^J~,
&
\big\langle \mathcal{O}_J(x)\,\overline{\mathcal{O}}_{J'}(0) \big\rangle_{\rm SYM} &= \frac{\delta_{JJ'}2^J}{\vert x\vert^{2J}}~,
\end{align}
where $u$ is a null polarization R-symmetry vector and we fix the normalization by specifying the two-point function in standard $\mathcal N=4$ SYM.
The operator naturally associated with the Lagrangian, $\mathcal O_{\mathcal L}$, is the top component obtained as the fourth supersymmetric descendant of $\mathcal O_2$ \cite{Dolan:2002zh}.
This is the operator that is holographically dual to the Type IIB dilaton in asymptotically $AdS_5\times S^5$ backgrounds. 
We will generally take $\mathcal O_{\mathcal L}$ to include the full coupling dependence, so that
\begin{align}
	S^{}_\text{$\mathcal N=4$ SYM}=\int d^4x\, \mathcal O_{\mathcal L}~.
\end{align}
$\mathcal O_{\mathcal L}$ differs from the standard form of the 4d $\mathcal N=4$ SYM Lagrangian by total derivative terms which change the form of the kinetic term for the scalars. 
This was spelled out for the Abelian theory in \cite{Basu:2004nt}, where
\begin{align}
	\mathcal O_{\mathcal L}&=\frac{i}{4\tau_2}\left(\tau \mathcal O_\tau-\bar\tau \bar {\mathcal O}_\tau\right)\,, &
	\tau&=\tau_1+i\tau_2=\frac{\theta}{2\pi}+\frac{4\pi i}{g_{\rm YM}^2}\,,
\end{align}
and $\mathcal O_\tau$ is the operator in the stress tensor multiplet given for the Abelian theory by
\begin{align}
	\mathcal O_{\tau}^{\rm Abelian}&=\frac{\tau_2}{8\pi}\left(F_{\mu\nu}F^{\mu\nu}+iF_{\mu\nu}\tilde F^{\mu\nu}-4i\bar\lambda\bar\sigma_\mu\partial^\mu\lambda^i+4\bar\varphi_{ij}\partial_\mu\partial^\mu\varphi^{ij}\right).
\end{align}
The explicit expression in the Abelian theory thus is
\begin{align}
	\mathcal O_{\mathcal L}
	&=-\frac{1}{4g_{\rm YM}^{2}}\left(F_{\mu\nu}F^{\mu\nu}
	-2i\left(
	\bar\lambda_i\bar\sigma_\mu\partial^\mu\lambda^i
	+\lambda^i\sigma_\mu\partial^\mu\bar\lambda_i
	\right)
	+2
	\left(
	\bar\varphi_{ij}\partial_\mu\partial^\mu\varphi^{ij}
	+\varphi^{ij}\partial_\mu\partial^\mu\bar\varphi_{ij}
	\right)
	\right)
	\nonumber\\
	&\hphantom{=}-\frac{\theta}{32\pi^2}\left(F_{\mu\nu}\widetilde F^{\mu\nu}
	-2\left(
	\bar\lambda_i\bar\sigma_\mu\partial^\mu\lambda^i
	-\lambda^i\sigma_\mu\partial^\mu\bar\lambda_i
	\right)
	-2i
	\left(
	\bar\varphi_{ij}\partial_\mu\partial^\mu\varphi^{ij}
	-\varphi^{ij}\partial_\mu\partial^\mu\bar\varphi_{ij}
	\right)\right).
\end{align}
The derivative structure was also noted in \cite{Clark:2004sb}. With boundaries, defects or interfaces, the action in general acquires additional terms localized on the codimension-1 feature, and for Janus interfaces we will actually allow $\tau$ to vary smoothly over an extended spacetime region. The object we refer to as ambient Lagrangian, however, remains $\mathcal O_{\mathcal L}$.
We note that $\mathcal O_{\mathcal L}$ and $\mathcal O_2$ being in the same superconformal multiplet in standard $\mathcal N=4$ SYM does not immediately imply a relation between their one-point functions in the presence of boundaries, defects or interfaces. The defects break at least half of the ambient superconformal symmetry, and the one-point functions are constrained only by the remaining symmetry. This means different components of an ambient multiplet may behave differently. The stress tensor, for example, belongs to the same 4d $\mathcal{N}=4$ multiplet as $\mathcal O_{\mathcal L}$ and $\mathcal O_2$ but has vanishing one-point function in planar codimension-one setups as a consequence of the defect conformal symmetry  \cite{McAvity:1995zd,Billo:2016cpy}.

A further quantity which will be relevant is the defect free energy. While the free energy of $\mathcal N=4$ SYM is scheme-dependent, suitable differences between the free energy with boundary or interface and the free energy of standard $\mathcal N=4$ SYM are scheme-independent.
We denote this quantity uniformly by $\mathcal F_{\rm def}$, with the following definitions: \bea \label{eq: def-free-en-gen}
\mathcal F_{\rm def}
\equiv
\begin{cases}
	\displaystyle
	\mathcal F_{\rm BCFT}-\frac{1}{2}\mathcal F_{S^4}
	& \text{BCFT,} \\[10pt]
	\displaystyle
	\mathcal F_{\rm ICFT}
	-\frac{1}{2}\mathcal F_{S^4}^{L}
	-\frac{1}{2}\mathcal F_{S^4}^{R}
	& \text{ICFT.}
\end{cases}
\eea
Here $\mathcal F_{\rm BCFT}\equiv -\log \mathcal Z_{HS^4}$ and $\mathcal F_{\rm ICFT}\equiv -\log \mathcal Z_{S^4}^{\rm interface}$ denote the free energies in the presence, respectively, of a boundary and of an interface, while $\mathcal F_{S^4}^{L}$ and $\mathcal F_{S^4}^{R}$ are the free energies for the ambient theories on the two sides of the interface. The defect case is recovered from the interface setup by taking the two bulk theories, and hence their free energies, to be identical. 
As emphasized in \cite{Bason:2023bin}, the Kähler ambiguity of the sphere partition function is matched by the corresponding ambiguity of the hemisphere partition function, once the appropriate boundary completion is included, and therefore cancels in $\mathcal F_{\rm def}$.  

In the remainder of this section we first discuss the general relation between integrated one-point functions of the Lagrangian and the defect free energy, then the computation of local and integrated correlators using localization, and in the last part how $\langle \mathcal O_{\mathcal L}\rangle$ can be related to both $\langle \mathcal O_2\rangle$ and to a derivative of the defect free energy for boundary, defect and interface theories which preserve at least a quarter of the supersymmetries.

\subsection{Integrated one-point function}\label{sec:integrated-L}

The technique we use to extract the one-point function of the Lagrangian density is to place the BCFT on a hemisphere $HS^4$, with boundary conditions imposed on the equatorial $S^3$, and then differentiate the partition function with respect to the Yang-Mills coupling (for much of this subsection we treat defect and interface CFTs in the folded picture as BCFTs). This derivative brings down an integrated insertion of the ambient Yang-Mills Lagrangian, potentially with boundary contributions. Placing the theory on $HS^4$ is essential as it makes the partition function finite and accessible to supersymmetric localization. In this subsection, however, we first discuss $\mathcal F_{\rm def}$ and $\langle \mathcal O_{\rm \mathcal L}\rangle$  without using supersymmetry.

We work with field normalizations such that the coupling appears as overall factor in the action, and to make this explicit we set $\mathcal O_{\mathcal L}=g_{\rm YM}^{-2}\mathcal L_{\rm YM}$, where $\mathcal L_{\rm YM}$ has no explicit coupling dependence. The action for a BCFT on the hemisphere can then be written as 
\bea \label{eq:HS4-action}
S & = \frac{1}{g_{\rm YM}^2} \!\left( \int_{HS^4}\!d^4x\,\sqrt{g(x)}\, \mathcal L_{\rm YM} +\int_{S^3}\!d^3x\,\sqrt{\gamma(x)}\, \mathcal L^{\partial}_{\rm YM}  \right)  +S_{3d}~.
\eea
Here $g_{\mu\nu}$ denotes the metric on the hemisphere and $\gamma_{ij}$ is the induced metric on the equatorial $S^3$. $\mathcal L_{\rm YM}$ is the curved-space Lagrangian, and $\mathcal L_{\rm YM}^{\partial}$ denotes possible boundary or interface terms for the ambient fields. Such terms are typically required by supersymmetry.
The equatorial $S^3$ may in addition host local defect degrees of freedom described by a 3d action $S_{3d}$. The 3d degrees of freedom can interact with the ambient $\mathcal N=4$ SYM fields, e.g.\ when gauging a global symmetry of the 3d theory using the ambient gauge field, but in the conventions used here $S_{3d}$ does not carry explicit dependence on the 4d coupling.\footnote{In our conventions with an overall $1/g_{\rm YM}^2$ in front of the Lagrangian, QFT vertices are $O(g_{\rm YM}^0)$ while propagators are $O(g_{\rm YM}^2)$. In this convention the coupling between the 4d fields and the 3d Lagrangian $\mathcal L^{3d}$ does not contain an explicit Yang-Mills coupling. This is reflected in the localized matrix model: the explicit dependence on the ambient ’t Hooft coupling is carried by the classical Gaussian factor, while the boundary and interface contributions enter through $Z_{\rm 1-loop}$, which has no explicit dependence on $\lambda$.}

We can now differentiate the defect free energy, $\mathcal F_{\rm def}$ in (\ref{eq: def-free-en-gen}), with respect to the gauge coupling $g_{\rm YM}$. Equivalently, at fixed $N$, one may differentiate the hemisphere partition function with respect to the 't Hooft coupling
\bea \label{eq: thooft}
\lambda \equiv g_{\rm YM}^2 N~.
\eea
We phrase the discussions in terms of $\lambda$, noting that $\lambda\partial_\lambda=g_{\rm YM}^2\partial_{g_{\rm YM}^2}$. The derivative brings down an insertion of the corresponding part of the action in $\mathcal Z_{HS^4}$ and $\mathcal Z_{S^4}$ and we obtain
\bea \label{eq:lambda-derivative-HS4}
\lambda \partial_\lambda \mathcal F_{\rm def}
=
-\frac{N}{\lambda}\left\langle\int_{HS^4} d^4x\,\sqrt{g(x)}\,
\mathcal L_{\rm YM}
+\int_{S^3} d^3x\,\sqrt{\gamma(x)}\,
\mathcal L^{\partial}_{\rm YM}\right\rangle_{HS^4, \ \rm subtracted},
\eea
where the subscript on the right hand side emphasizes that the one-point function on $S^4$ has been subtracted as a result of subtracting $\mathcal Z_{S^4}$ in $\mathcal F_{\rm def}$.

One-point functions on $(H)S^4$ differ from flat-space one-point functions not only in their transformed coordinate dependence, but they are also affected by mixing.
In order to extract the correct flat space one-point function $\mathcal{C}_{O_{\mathcal{L}_{\rm YM}}}$, one has to take Weyl anomalies into account \cite{Gerchkovitz:2016gxx}. When transforming to $(H)S^4$, local curvature invariants in general induce mixings between operators of different conformal dimensions. In the present case, the only relevant mixing is with the identity operator. This is removed by subtracting the one-point function on $S^4$, so that flat-space one-point functions can be extracted directly from \eqref{eq:lambda-derivative-HS4}.

\paragraph{Relation to $\langle\mathcal O_{\mathcal L}\rangle$:}
The goal is to extract a local one-point function from the integrated correlator in \eqref{eq:lambda-derivative-HS4}. The formal expression for the integrated correlator is divergent and must be regularized. After adding suitable counterterms, its finite part is well-defined and determines the coefficient of the flat-space one-point function. This finite part is in turn related to the (finite) derivative of $\mathcal F_{\rm def}$ with respect to the 't Hooft coupling via \eqref{eq:lambda-derivative-HS4}.

To make this discussion more explicit, we take the metric for the round hemisphere of unit radius as
\bea \label{eq: metricHS}
ds^2_{HS^4} = d\theta^2 + \sin^2\!\theta \,d\Omega_3^2~, \qquad 0\leq\theta\leq \frac{\pi}{2}~.
\eea
The equator, which may host boundary or defect degrees of freedom, is located at $\theta=\frac{\pi}{2}$. Under the Weyl map from the flat half-space to the hemisphere, the flat-space distance from the boundary is mapped to $\cos\theta$, and one-point functions on $HS^4$ take the form
\bea \label{eq: one-point-gen-HS}
\expval{\mathcal O_\Delta}_{\!HS^4}=\frac{\mathcal{C}_{\mathcal O}}{\left(\cos\theta\right)^{\Delta_{\mathcal O}}}~.
\eea
Using the metric \eqref{eq: metricHS} and the equation \eqref{eq: one-point-gen-HS}, the ambient integral in \eqref{eq:lambda-derivative-HS4} becomes
\bea \label{eq: integral-on-ads}
\int_{HS^4} d^4x\,\sqrt{g(x)}\,
\big\langle \mathcal L_{\rm YM}\big\rangle_{\!HS^4, \rm subtracted} =2\,\pi^2\,\mathcal{C}_{\mathcal O_{\mathcal{L}_{\rm YM}}}\int_{0}^{\frac{\pi}{2}} \!d\theta\,
\frac{\sin^3\theta}{\cos^4\theta}~.
\eea
As discussed above, the subtracted $HS^4$ one-point function directly gives $\mathcal{C}_{\mathcal O_{\mathcal{L}_{\rm YM}}}$ on flat space.
This integral is divergent, but can be regulated by introducing a cut-off $\delta$,
\bea \label{eq: int-HS}
2\,\pi^2\,\int_{0}^{\frac{\pi}{2}-\delta} \!d\theta\,
\frac{\sin^3\theta}{\cos^4\theta} =
\frac{2\,\pi^2}{3\delta^3}-\frac{5\pi^2}{3\delta}+\frac{4\pi^2}{3}+ O\left( \delta \right) ~.
\eea
The power divergences may be removed by local counterterms \cite{Herzog:2019bom}, leaving the finite term as the physical contribution relevant for extracting the one-point coefficient.\footnote{The volume $4\pi^2/3$ matches the renormalized volume of unit-radius $AdS_4$ relevant in holography.} Combining \eqref{eq:lambda-derivative-HS4}, \eqref{eq: integral-on-ads} and \eqref{eq: int-HS}, we find
\bea \label{eq: der-one-point-HS}
\lambda \partial_\lambda \mathcal F_{\rm def} = -\frac{4\pi^2}{3g_{\rm YM}^2}\,\mathcal C_{\mathcal O_{\mathcal L_{\rm YM}}}  + \text{boundary contributions}~,
\eea
where the boundary contributions are the terms in \eqref{eq:lambda-derivative-HS4} localized on $S^3$.

The above analysis uses the derivative with respect to the coupling on one half space. It also applies to setups with independent couplings on two half spaces. For a two-sided ambient theory with fixed overall coupling, both sides of the defect contribute to the integrated correlator, so that the geometric factor is doubled. The final relation becomes
\bea \label{eq: Lfin2}
\lambda\partial_\lambda \mathcal F_{\rm def}
=
-\frac{8\pi^2}{3g_{\rm YM}^2}\mathcal C_{\mathcal O_{\mathcal L_{\rm YM}}}
\,+ \text{defect contributions}~.
\eea

It remains to discuss the contributions localized on the boundary/defect. While one-point functions of non-identity conformal primaries localized on the codimension-1 feature vanish by conformal invariance in flat space, after the Weyl map to the sphere a 3d primary operator localized on $S^3$ could in principle contribute through mixing with the identity. Such mixing is absent on odd-dimensional spheres \cite{Gerchkovitz:2014gta}. If the boundary terms are primaries, $\lambda\partial_\lambda\mathcal F_{\rm def}$ therefore directly gives the one-point function $\mathcal C_{\mathcal O_{\mathcal L_{\rm YM}}}$. Complications are that the split into ambient Lagrangian and boundary terms is ambiguous due to integration-by-parts identities, and that Janus interfaces can exhibit distributional boundary terms \cite{DHoker:2006qeo} and coupling-dependent gluing conditions, which call for the interface to be viewed as limit of theories with smooth coupling profiles. We will sidestep these issues and relate $\lambda\partial_\lambda\mathcal F_{\rm def}$ to the local one-point function $\langle\mathcal O_{\mathcal L}\rangle$ using supersymmetry in the next two subsections.

\subsection{One-point functions from localization}

We now discuss the relevant aspects of supersymmetric localization, which will be used for the computations below. For localization to be applicable we place the theory on $S^4$, with boundary, defect or interface along an equatorial $S^3$. One picks a supercharge $Q$, and adds a $Q$-exact term to the action which suppresses all but a finite-dimensional space of contributions, leading to a matrix model. This permits the exact computation of the partition function and insertions that are invariant under $Q$: for an insertion into the path integral to be captured by the matrix model, it needs to be invariant under $Q$ and $Q^2$, where, schematically, $Q^2\sim \text{gauge transformation} + \text{isometry} + \text{R-symmetry transformation}$.
When it comes to correlators there are two strategies: either directly evaluate local insertions on $(H)S^4$, or compute integrated correlators. For the situation at hand they matter as follows:
\begin{itemize}
	\item[--] The chiral primary $\mathcal O_2$ is the bottom component in its multiplet; at a pole on $S^4$ and for a suitable choice of R-symmetry polarization, $\mathcal O_2$ is invariant under $Q^2$ and $Q$. It can therefore be evaluated as a local operator insertion on $(H)S^4$ using localization.
	
	\item[--] The ambient Lagrangian $\mathcal O_{\mathcal L}$ transforms into a total derivative under $Q$; only integrated versions can be $Q$-invariant and need suitable interface terms $\mathcal O_\partial$. We can evaluate $\int\langle\mathcal O_{\mathcal L}+\mathcal O_\partial\rangle$ using localization, but one still needs to separate the local one-point function $\langle\mathcal O_{\mathcal L}\rangle$ from potential interface-localized contributions to the integral.
\end{itemize}
Chiral primary operator expectation values have been computed directly as local insertions e.g.\ in \cite{Komatsu:2020sup,Bason:2023bin,He:2025due}. Here we review the mechanism in connection with the coupling-derivative of the defect free energy.

We start with a BCFT; the generalization to defects and interfaces will be discussed shortly. Localization reduces the path integral to a finite-dimensional matrix model, whose variables include the eigenvalues associated with the four-dimensional ambient gauge group and possibly variables associated with boundary degrees of freedom. Schematically, the resulting matrix model can be written as 
\bea
\label{eq:BCFT-matrix-model-gen}
Z_{HS^4}
=
\int
\prod_{i=1}^{N} da\,\prod_{i<j}\left(a_{i}-a_{j}\right)^2
\exp\!\left[
-\frac{4\pi^2 N}{\lambda}
\sum_{i=1}^{N} a_{i}^2
\right]\,
Z_{\rm 3d}\!\left(\{a\}\right)~.
\eea
We made the 4d part explicit, with eigenvalues $a_{i}$. The Gaussian factor is the classical action of 4d $\mathcal N=4$ SYM on $HS^4$, written in terms of the ambient 't Hooft coupling.\footnote{The same form arises for the more general class of theories discussed in section \ref{sec:O2-OL}.} $Z_{\rm 3d}$ denotes additional contributions associated with the boundary, e.g.\ in the form of additional matrix integrals, one-loop determinants etc.\ associated with 3d degrees of freedom. The explicit form depends on the BCFT under consideration; the relevant feature for our purposes is that $Z_{\rm 3d}$ does not have explicit dependence on the 4d marginal coupling $\lambda$.

Using the definition of the boundary free energy \eqref{eq: def-free-en-gen}, together with the matrix model for $\mathcal{N}=4$ SYM on $S^4$,
\begin{align} \label{eq: SYM-matrix}
	\mathcal Z_{S^4}&=\int
	\prod_{i=1}^{N} da_{i}\,\prod_{i<j}(a_i-a_j)^2 \exp\!\left[-\frac{8\pi^2 N}{\lambda}\sum_{i=1}^{N} a_{i}^2\right]~,
\end{align}
we find
\begin{align}\label{eq:dFdlambdaBCFT}
	\lambda\,\partial_{\lambda}\mathcal F_{\rm def}=-\frac{4\pi^2N}{\lambda}\left( \big\langle {\rm tr}\, a^2 \big\rangle -\big\langle {\rm tr}\, a^2 \big\rangle_0 \right) ~.
\end{align}
Here $\big\langle\cdot\big\rangle_0$ denotes expectation values in the pure Gaussian matrix model on $S^4$.
This connects to the chiral primary operator $\mathcal O_2$ due to the fact that, on the localization locus, $\mathcal O_2$ reduces precisely to ${\rm tr}\,a^2$. The combination appearing in \eqref{eq:dFdlambdaBCFT} therefore also computes the one-point function of $\mathcal O_2$. The subtraction of $\big\langle {\rm tr}\,a^2 \big\rangle_0$ implements the normal-ordering prescription which disentangles the operator mixing incurred when mapping from flat space to the sphere: the only operator $\mathcal O_2$ can mix with on $HS^4$ is the identity, whose contribution is fixed by the Gaussian matrix model. In particular, using \cite[(A.9)]{He:2025due}, we have
\begin{align}
	\mathcal{C}_{\mathcal{O}_2}=-\frac{4\sqrt{2}\,\pi^2}{\lambda}\left( \big\langle {\rm tr}\, a^2 \big\rangle -\big\langle {\rm tr}\, a^2 \big\rangle_0 \right) =\frac{\sqrt{2}}{N}\lambda\partial_\lambda \mathcal F_{\rm def}~.
	\label{eq: O2normord}
\end{align}
This establishes a relation between the coupling-derivative of the defect free energy and $\langle \mathcal O_2\rangle$ as a local one-point function. \footnote{This relation can also be recovered from \cite[(3.17)]{Bason:2023bin}, which states $a_A(\tau,\bar\tau)=\frac{1}{2\pi i}\partial_\tau\mathcal F_{\rm def}$, noting that $a_A(\tau,\bar\tau)$ is related to our $\mathcal O_2$ by $a_A(\tau,\bar\tau)= \frac{\lambda}{16\sqrt{2}\,\pi^2}\,\langle\mathcal O_2\rangle$ and that $\langle \mathcal O_2\rangle = \langle \overline{\mathcal O}_2\rangle$ when $\theta=0$.}

We note that the action also reduces to $\tr a^2$ on the localization locus and can thus be related to $\lambda\partial_\lambda \mathcal F_{\rm def}$. However, supersymmetry requires the insertion to be the integrated Lagrangian with all necessary boundary terms, and the comments at the end of section \ref{sec:integrated-L} regarding the extraction of a local operator insertion of $\mathcal O_{\mathcal L}$ apply here as well.

The extension of (\ref{eq: O2normord}) to interfaces and defects is immediate. One works on the full $S^4$ with interface or defect on an equatorial $S^3$. For interfaces with independent 4d couplings on the two half spaces, (\ref{eq:BCFT-matrix-model-gen}) contains a second 4d $\mathcal N=4$ SYM node with independent coupling. Its contribution drops out in the derivative in (\ref{eq:dFdlambdaBCFT}), leaving (\ref{eq: O2normord}) unchanged. For defects the coupling is $\lambda$ on the entire $S^4$. This doubles the geometric volume factor in (\ref{eq:dFdlambdaBCFT}) and in turn leads to a factor two in \eqref{eq: O2normord}. In summary,
\begin{align}
	\text{B/ICFT:\quad} \mathcal{C}_{\mathcal{O}_2} &=\frac{\sqrt{2}}{N}\lambda\partial_\lambda \mathcal F_{\rm def}~,
	&
	\text{dCFT:\quad} \mathcal{C}_{\mathcal{O}_2} &=\frac{1}{\sqrt{2}N}\lambda\partial_\lambda \mathcal F_{\rm def}~.
	\label{eq: O2normord-BIdCFT}
\end{align}

\smallskip

\subsection{\texorpdfstring{$\langle \mathcal O_{\mathcal L}\rangle$}{} from \texorpdfstring{$\langle \mathcal O_2\rangle$}{} with 3d \texorpdfstring{$\mathcal N=2$}{} supersymmetry}\label{sec:O2-OL}

We now derive a relation between $\langle \mathcal O_{\mathcal L}\rangle$ and $\langle \mathcal O_2\rangle$ as one-point functions of local operators using supersymmetry. We work in the context of 4d $\mathcal N=2$ SYM coupled to hypermultiplets, enriched by an arbitrary complex coupling profile depending on one spatial direction,
\begin{equation}
\tau(x)=\frac{\theta(x)}{2\pi}+\frac{4\pi i}{g_{\rm YM}(x)^2}~,
\end{equation}
while preserving half the supersymmetries which form a 3d $\mathcal N=2$ algebra. This is also the symmetry preserved by $\tfrac{1}{2}$-BPS boundaries, defects, and interfaces in 4d $\mathcal N=2$ SYM. The setup allows us to capture Janus interfaces, which correspond to step-function profiles for the coupling and can be subtle due to distributional interface terms, in a controlled way as limit of theories with smooth coupling profiles. At the same time, it captures general boundary, defect and interface theories preserving 3d $\mathcal N=2$ supersymmetry as long as the operators $\mathcal O_{\mathcal L}$ and $\mathcal O_2$ are in the ambient space. Conformal symmetry will not be used.

\paragraph{Review of the Lagrangian:}
We follow the discussion in \cite{Goto:2020per} and start with the 4d $\mathcal N=2$ vector multiplet in flat space, where we seek to relate the one-point functions. The addition of hypermultiplets, e.g.\ to extend the ambient theory to $\mathcal N=4$ SYM, will be discussed shortly.
The key to preserving supersymmetry is to embed the complex coupling $\tau$ into a chiral multiplet $\mathcal T$ of Weyl weight zero and its conjugate $\bar{\mathcal T}$, where
\begin{align}
	\mathcal{T}&=(\tau,\Psi^{(\tau)}_{ i},B_{ij}^{(\tau)},F_{\mu\nu}^{(\tau)-},\Lambda_{ i}^{(\tau)}, C^{(\tau)}) \ .
\end{align}
The fermions and $F_{\mu\nu}^{(\tau)\pm}$ are set to zero. To preserve 3d $\mathcal N=2$ supersymmetry, half of the fermion supersymmetry variations are required to vanish. The fermion variations are
\begin{align}
	\delta\Psi_{i}^{(\tau)}&=
	(\partial_3 \tau)  \gamma^3   \epsilon_{i}+\frac{1}{2}\,B^{(\tau)}_{ij}\epsilon^{j} \ ,
	&
	\delta \Lambda_{i}^{(\tau)}&= -\frac12 \partial_3 B_{ij}^{(\tau)}  \varepsilon^{jk} \gamma^3 \epsilon_k + \frac12 C^{(\tau)} \varepsilon_{ij} \epsilon^j \ ,
\end{align}
and the solution preserving half the supersymmetry is
\begin{align}\label{eps-N2susy}
	\epsilon_i &= \rho_{ij} \gamma^3 \epsilon^j \ , & 
	B^{(\tau)}_{ij} &= -2  \rho_{ij} \,\partial_3\tau \ , &
	C^{(\tau)}  &= -2 e^{+2{\rm i}\alpha}\, \partial_3^2\tau \ , 
\end{align}
where, with a phase $\alpha$ and real unit vector $\vec{n}$,
\begin{align} \label{rho-alpha-n}
	\rho_{ij} &= e^{{\rm i}\alpha}\,\vec n \cdot\vec\tau_{ij}  \ .
\end{align}
Here $\vec\tau_i^{\hphantom{i}j}=i\vec\sigma_i^{\hphantom{i}j}$, with $\vec\sigma$ denoting the Pauli matrices, and indices are raised/lowered with $\varepsilon^{ij}$, $\varepsilon_{ij}$.
The additional fields in the coupling multiplet vanish for constant coupling and encode the couplings of interface terms required for supersymmetry.

To construct the Lagrangian, the vector multiplet $\mathcal V=(X,\Omega_i,A_\mu,Y_{ij})$ is embedded in a chiral multiplet
$\mathcal{A}(\mathcal{V})$  of weight $w=1$, with components $(A,\Psi_i,B_{ij},F_{ab}^-,\Lambda_i,C)$, via
\begin{align}
	A|_{\mathcal{A}({\mathcal{V}})} &=X\,,
	&
	\Psi_{i}|_{\mathcal{A}({\mathcal{V}})} &=\Omega_{i}\,,
	&
	B_{ij}|_{\mathcal{A}({\mathcal{V}})}&=Y_{ij}\ ,
	\nonumber\\
	F_{ab}^{-}|_{\mathcal{A}({\mathcal{V}})} &=
	\frac12 \left(F_{ab}-\tilde{F}_{ab}\right), &
	\Lambda_{i}|_{\mathcal{A}({\mathcal{V}})} &=-\varepsilon_{ij}\,\gamma^\mu D_\mu\Omega^{j}\,,
	&
	C|_{\mathcal{A}({\mathcal{V}})} &=-2D_\mu D^\mu \overline{X}\,.
\end{align}
The Lagrangian is given by the $C$ components of the product multiplet $\mathcal{T}\mathcal{A}({\mathcal{V}})^{2}$, whose components can be constructed following the summary in \cite{Goto:2020per}, by
\begin{align}\label{eq:N2action}
	S_\text{$\mathcal N=2$ SYM, $\tau(x)$}
	=\frac{1}{8\pi i}\int d^{4}x\sqrt{g}\,
	{\rm tr} \left[C|_{\mathcal{T}\mathcal{A}({\mathcal{V}})^{2}}
	-
	\bar{C}|_{\bar{\mathcal{T}}\,\bar{\mathcal{A}}({\mathcal{V}}){}^{2}}
	\right] .
\end{align}

\paragraph{Relating one-point functions:} The product multiplet $\mathcal{A}({\mathcal{V}})^{2}$ to which $\mathcal T$ couples is a chiral multiplet of weight 2 and the supersymmetry transformations can be found in \cite[(A.25)--(A.30)]{Goto:2020per}. Taking the expectation value yields
\begin{align}
	\left\langle\delta\Psi_{i}\vert_{\mathcal{A}({\mathcal{V}})^{2}}\right\rangle&=
	\partial_3\left\langle A\vert_{\mathcal{A}({\mathcal{V}})^{2}}\right\rangle  \gamma^3   \epsilon_{i}+\frac{1}{2}\,\left\langle B_{ij}\vert_{\mathcal{A}({\mathcal{V}})^{2}}\right\rangle\epsilon^{j} \ ,
	\nonumber\\
	\left\langle \delta \Lambda_{i}\vert_{\mathcal{A}({\mathcal{V}})^{2}}\right\rangle&= -\frac12 \partial_3 \left\langle B_{ij}\vert_{\mathcal{A}({\mathcal{V}})^{2}}\right\rangle  \varepsilon^{jk} \gamma^3 \epsilon_k + \frac12 \left\langle C\vert_{\mathcal{A}({\mathcal{V}})^{2}} \right\rangle  \varepsilon_{ij} \epsilon^j \ .
\end{align}
For these expressions we used that $\left\langle F_{ab}^\pm\vert_{\mathcal{A}({\mathcal{V}})^{2}} \right\rangle=0$ due to  transverse Lorentz invariance, and we note that fermionic one-point functions vanish. Imposing that the left hand sides vanish and using the preserved supersymmetries $\epsilon_i$ and $\epsilon^i$ in (\ref{eps-N2susy}) leads to
\begin{align}
	0&=
	\partial_3\left\langle A\vert_{\mathcal{A}({\mathcal{V}})^{2}}\right\rangle   \rho_{ij} \epsilon^{j}+\frac{1}{2}\,\left\langle B_{ij}\vert_{\mathcal{A}({\mathcal{V}})^{2}}\right\rangle\epsilon^{j} \ ,
	\nonumber\\
	0&= -\frac12 \partial_3 \left\langle B_{ij}\vert_{\mathcal{A}({\mathcal{V}})^{2}}\right\rangle  \varepsilon^{jk}\rho_{kl} \epsilon^l + \frac12 \left\langle C\vert_{\mathcal{A}({\mathcal{V}})^{2}} \right\rangle  \varepsilon_{ij} \epsilon^j \ .
\end{align}
Combining the two equations leads to the condition
\begin{align}
	0&= \partial_3^2  \left\langle A\vert_{\mathcal{A}({\mathcal{V}})^{2}}\right\rangle \rho_{ij}\varepsilon^{jk}\rho_{kl} \epsilon^l + \frac12 \left\langle C\vert_{\mathcal{A}({\mathcal{V}})^{2}} \right\rangle  \varepsilon_{ij} \epsilon^j \ .
\end{align}
Expressing $\rho$ in terms of $\vec{\tau}=i\vec{\sigma}$ leads to
\begin{align}\label{eq:A-to-C-relation}
	\left\langle C\vert_{\mathcal{A}({\mathcal{V}})^{2}} \right\rangle&=
	-2e^{2i\alpha}\partial_3^2  \left\langle A\vert_{\mathcal{A}({\mathcal{V}})^{2}}\right\rangle  \ .
\end{align}
The bottom component is $A\vert_{\mathcal{A}({\mathcal{V}})^{2}}=X^2$, which means upon taking the trace we get the chiral primary. The object appearing in the Lagrangian is $C\vert_{\mathcal{T}\mathcal{A}({\mathcal{V}})^{2}}$, which is given by
\begin{align}
	C|_{\mathcal{T}\mathcal{A}({\mathcal{V}})^{2}}
	&=
	A^{(\tau)}\, C|_{\mathcal{A}({\mathcal{V}})^{2}}
	+
	C^{(\tau)}\, A|_{\mathcal{A}({\mathcal{V}})^{2}}
	-\frac{1}{2}\,\varepsilon^{ik}\varepsilon^{jl}B_{ij}^{(\tau)}\, B_{kl}|_{\mathcal{A}({\mathcal{V}})^{2}}\,.
\end{align}
The component multiplying the complex coupling $A^{(\tau)}=\tau$ in the action (\ref{eq:N2action}) is $C|_{\mathcal{A}({\mathcal{V}})^{2}}$; the other components couple to derivatives of $\tau$ and are interface terms which localize in the sharp-interface limit. So $C|_{\mathcal{A}({\mathcal{V}})^{2}}$ is the holomorphic part of the ambient Lagrangian,
\begin{align}\label{eq: c-to-L}
	\mathcal O_{\mathcal L}&=\frac{1}{8\pi i}\tr\left(\tau C|_{\mathcal{A}({\mathcal{V}})^{2}} - \bar\tau \bar C|_{\mathcal{A}({\mathcal{V}})^{2}}\right)~.
\end{align}
Equation (\ref{eq:A-to-C-relation}) and its conjugate therefore express the one-point function of the ambient Lagrangian in terms of the one-point function of the chiral primary,
\begin{align}\label{eq:A-to-OL-rel}
	\langle\mathcal O_{\mathcal L}\rangle&=-\frac{1}{4\pi i}\partial_3^2\left(\tau \langle\tr e^{2i\alpha}X^2\rangle-\bar\tau \langle\tr e^{-2i\alpha} \bar X^2\rangle\right)~.
\end{align}

We note that we can write $\mathcal O_{\mathcal L}$ as variation of the action in this formulation with respect to the $A^{(\tau)}$ component of the coupling multiplet while keeping the other components fixed,
\begin{align}\label{eq:dS-OL}
	\mathcal O_{\mathcal L}&=\left(\tau\frac{\delta}{\delta\tau}+\bar\tau\frac{\delta}{\delta\bar\tau}\right)S_\text{$\mathcal N=2$ SYM, $\tau(x)$}\Big\vert_\text{$C^{(\tau)},\overline{C}^{(\tau)},B_{ij}^{(\tau)},B^{(\tau)ij}$ fixed}\,.
\end{align}
The functional derivative directly extracts a local quantity from the action. For this to be defined, the formulation of the theory in a general supergravity background is crucial.
We note that this is also natural from the holographic perspective, where arbitrary position-dependent boundary values can be imposed for the dilaton and axion.

\paragraph{Adding hypermultiplets:} We now incorporate hypermultiplets in an arbitrary representation of the gauge group, denoted $\mathcal H$. For a single adjoint hypermultiplet this upgrades the ambient theory from 4d $\mathcal N=2$ SYM to 4d $\mathcal N=4$ SYM. Since hypermultiplets do not couple directly to the coupling multiplet $\mathcal T$, and only notice the coupling profile through their coupling to the vector multiplet $\mathcal V$, they incorporate straightforwardly into the constructions of \cite{Goto:2020per}. The action takes the form
\begin{align}
	S_\text{$\mathcal N=4$ SYM, $\tau(x)$}&=S_H[\mathcal H,\mathcal V]+\frac{1}{8\pi i}\int d^{4}x\sqrt{g}\,
	{\rm tr} \left[C|_{\mathcal{T}\mathcal{A}({\mathcal{V}})^{2}}
	-
	\bar{C}|_{\bar{\mathcal{T}}\,\bar{\mathcal{A}}({\mathcal{V}}){}^{2}}
	\right] ,
\end{align}
where the first term encodes the hypermultiplet and its coupling to $\mathcal V$.
The components $C|_{\mathcal{T}\mathcal{A}({\mathcal{V}})^{2}}$ and $\bar{C}|_{\bar{\mathcal{T}}\,\bar{\mathcal{A}}({\mathcal{V}}){}^{2}}$ pick up hypermultiplet contributions indirectly (the auxiliary fields in $\mathcal A(\mathcal V)$, which enter $C_{\mathcal A(\mathcal V)^2}$, capture hypermultiplet contributions on shell due to their coupling to the hypermultiplet), but do not include e.g.\ the hypermultiplet kinetic terms. We define a partial ambient Lagrangian operator as before as
\begin{align}
	{\mathcal O}^0_{\mathcal L}&=
	\frac{1}{8\pi i}\tr\left(\tau C|_{\mathcal{A}({\mathcal{V}})^{2}} - \bar\tau \bar C|_{\bar{\mathcal{A}}({\mathcal{V}})^{2}}\right)
	\,.
\end{align}
The relation (\ref{eq:A-to-C-relation}) only relied on the preserved supersymmetry and the structure of the vector multiplet. It remains valid and now fixes $\langle{\mathcal O}^0_{\mathcal L}\rangle $ in terms of the one-point function of the chiral primary,
\begin{align}\label{eq:A-to-OL0-rel}
	\langle\mathcal O_{\mathcal L}^0\rangle&=-\frac{1}{4\pi i}\partial_3^2\left(\tau \langle\tr e^{2i\alpha}X^2\rangle-\bar\tau \langle\tr e^{-2i\alpha} \bar X^2\rangle\right)~.
\end{align}

In order to relate $\langle\mathcal O_{\mathcal L}^0\rangle$ to the one-point function of the full ambient Lagrangian, we note that $\mathcal O_{\mathcal L}$ and $\mathcal O_{\mathcal L}^0$ differ only in the hypermultiplet terms, and that the hypermultiplet action is quadratic in the component fields before eliminating auxiliary fields, $S_H[t\mathcal H,\mathcal V]=t^2S_H[\mathcal H,\mathcal V]$.
This can be used to extract the Yang-Mills coupling as overall factor, and implies that a representative for the full ambient Lagrangian can be obtained as
\begin{align}\label{eq:S-Stilde-rel}
	\mathcal O_{\mathcal L}&=\mathcal O_{\mathcal L}^0+\frac{1}{2}\phi^i_{\mathcal H}\frac{\delta S}{\delta\phi_{\mathcal H}^i}~,
\end{align}
where $\phi_{\mathcal H}^i$ denotes the hypermultiplet component fields with Lorentz indices suppressed. We note that the kinetic term for the hypermultiplet scalar fields in (\ref{eq:S-Stilde-rel}) takes the form $\phi\square\phi$.
In the one-point function, the second term in (\ref{eq:S-Stilde-rel}) can be evaluated using the Schwinger-Dyson equation $\langle F \delta S/\delta \phi\rangle=\langle \delta F/\delta\phi\rangle$ and point splitting, which gives contributions of the form $\lim_{y\rightarrow x}\delta(y-x)$. These do not change the coefficient of $1/|x_3|^\Delta$.\footnote{E.g.\ in dimensional regularization $\delta^{(d)}(0)=\int d^dp/(2\pi)^d=0$.} 
We conclude that the one-point function coefficients for $\langle\mathcal O_{\mathcal L}\rangle$ and $\langle\mathcal O_{\mathcal L}^0\rangle$ in the notation of (\ref{eq: one-point-gen}) satisfy
\begin{align}\label{eq:CL-CL0}
	\mathcal C_{\mathcal O_{\mathcal L}}&=\mathcal C_{\mathcal O_{\mathcal L}^0}~.
\end{align}
This extends the relation between $\langle \mathcal O_{\mathcal L}^0\rangle$ and $\langle \mathcal O_2\rangle$ in (\ref{eq:A-to-OL0-rel}) to a relation between the one-point functions of the chiral primary and of the full ambient Lagrangian operator.

\paragraph{Application to $\mathcal N=4$ SYM:} Adding an adjoint hypermultiplet extends the ambient theory to 4d $\mathcal N=4$ SYM. The coupling-variation-dependent terms, however, remain tied to the 4d $\mathcal N=2$ vector multiplet. With general coupling profiles the preserved supersymmetry remains 3d $\mathcal N=2$, with an $SU(2)$ flavor symmetry acting on the adjoint hypermultiplet scalars. Other $\mathcal N=2$ Janus variants as well as the maximal $\mathcal N=4$ Janus theory differ only in interface couplings \cite{DHoker:2006qeo}. These do not enter the Lagrangian away from the interface and do not change the relation between $\langle\mathcal O_{\mathcal L}\rangle$ and $\langle\mathcal O_2\rangle$. The same applies to introducing boundaries, defects, or more general interfaces as long as they do not break additional symmetries: the local analysis in the ambient space carries over. 

Matching the normalization of the scalar $X$ with that of the chiral primary defined in (\ref{eq:OJ-gen}) leads to the identification $\mathcal O_2=\frac{16\sqrt{2}\pi^2}{\lambda}\tr e^{2i\alpha}X^2$. Therefore, the relation between the one-point function coefficients, using (\ref{eq:A-to-OL0-rel}) and (\ref{eq:CL-CL0}), becomes
\begin{align}
	\mathcal C_{\mathcal O_{\mathcal L}}&=-\frac{3\sqrt{2}\lambda}{8\pi^2}\frac{1}{8\pi i}\left(\tau\mathcal C_{\mathcal O_2}-\bar\tau\mathcal C_{\overline{\mathcal O}_2}\right)~.
\end{align}
For $\theta=0$, when $\langle \mathcal O_2\rangle=\langle\bar{\mathcal O}_2\rangle$, this simplifies to
\begin{align}\label{eq:COL-CO2}
	\mathcal C_{\mathcal O_{\mathcal L}}&=-\frac{3\sqrt{2}N}{8\pi^2}\mathcal C_{\mathcal O_2}~.
\end{align}
With (\ref{eq:COL-CO2}), the relations in (\ref{eq: O2normord-BIdCFT}) also determine the local ambient Lagrangian one-point function $\langle \mathcal O_{\mathcal L}\rangle$ in terms of the defect free energy, as\footnote{%
Comparing to (\ref{eq: der-one-point-HS}), (\ref{eq: Lfin2}), we note that ignoring boundary/defect contributions there gives the same relations. This derivation, though, explains for which form of the Lagrangian the relations hold.}
\begin{align} \label{eq: dF-to-L}
	\text{B/ICFT:\quad} \mathcal{C}_{\mathcal O_{\mathcal L}}&=-\frac{3}{4\,\pi^2} \lambda\,\partial_{\lambda}\mathcal F_{\rm def} ~,
	&
	\text{dCFT:\quad} \mathcal{C}_{\mathcal O_{\mathcal L}}&=-\frac{3}{8\,\pi^2} \lambda\,\partial_{\lambda}\mathcal F_{\rm def} ~,
\end{align}
where the derivatives should be understood with respect to the coupling at the pole.
The relations in \eqref{eq: dF-to-L} play a central role in our analysis. The boundary or interface free energy can be computed exactly by supersymmetric localization, in some cases for finite values of the coupling $\lambda$ and the rank $N$. These relations therefore provide direct access to non-perturbative information about the Lagrangian one-point function.

\section{\texorpdfstring{$\man=4$}{} SYM with \texorpdfstring{$\frac{1}{2}$}{}-BPS boundaries, defects and interfaces}\label{sec: SYM}

$\man=4$ SYM with general $\frac{1}{2}$-BPS boundaries or interfaces was constructed in \cite{Gaiotto:2008sa,Gaiotto:2008ak}. We  briefly review it in the following.
The action of $U(N)$ $\man=4$ SYM on $\mathbb R^{1,3}=(x^0,x^1,x^2,x^3)$ is conveniently obtained from 10d $\man=1$ SYM by dimensional reduction \cite{Brink:1976bc}
\bea\label{eq: ac-sym}
S_{\text{SYM}}=-\frac{1}{\gym^2}\int d^4 x \tr (\frac{1}{4}F_{MN}F^{MN}-\Psi\Gamma^M D_M\Psi)~,
\eea
where the fermions are Majorana-Weyl type and transform as \textbf{16} of $SO(1,9)$.
The 10d index $M=0,1,\cdots,9$ splits into 4d spacetime $SO(1,3)$ and $SO(6)$ internal R-symmetry indices $(\mu=0,1,2,3;I=4,5,6,7,8,9)$.
$\Gamma_M$ is the 10d Gamma matrix, and the trace is taken with respect to the gauge group.
Upon dimensional reduction, the 10d vector splits into a 4d vector $A_{\mu}$ and 6 adjoint scalars $\Phi_I$.
The fermions transform as $(\mathbf{2},\mathbf{1},\mathbf{4})\oplus(\mathbf{1},\mathbf{2},\mathbf{\bar 4})$ under the remaining $SO(1,3)\times SO(6)$ symmetry.
The conformal symmetry and supersymmetry combine into the superconformal symmetry $PSU(2,2|4)$ with bosonic subgroup $SO(2,4)\times SO(6)$ including the 4d conformal symmetry and R-symmetry.

\begin{figure}
	\centering
	\begin{tikzpicture}[xscale=2.2,yscale=1.6]
		\foreach \i in {-0.1,0,0.1} \draw[thick] (-0.5,\i) -- (0,\i);
		\foreach \i in {-0.1,0,0.1} \draw[thick] (0,\i) -- (0.8,\i);
		\foreach \i in {-0.15,-0.05,0.05,0.15} \draw[thick] (0.8,\i) -- (1.6,\i);
		\foreach \i in {-0.2,-0.1,0,0.1,0.2} \draw[thick] (1.6,\i) -- (2.1,\i);
		
		\node at (2.5,0) {\ldots};	
		
		\foreach \i in {-0.2,-0.1,0,0.1,0.2} \draw[thick] (2.9,\i) -- (3.5,\i);
		\foreach \i in {-0.15,-0.05,0.05,0.15} \draw[thick] (3.5,\i) -- (4.3,\i);
		\foreach \i in {-0.1,0,0.1} \draw[thick] (4.3,\i) -- (4.8,\i);
		
		\foreach \i in {0,0.8,1.6,3.5,4.3}{ \draw[fill=gray] (\i,0) ellipse (2pt and 7pt);}
		
		\foreach \i in {-0.05,0.05} \draw (0.4+\i,-0.8) -- +(0,1.6);
		\foreach \i in {0} \draw (1.2+\i,-0.8) -- +(0,1.6);
		
		\foreach \i in {-0.075,0,0.075} \draw (3.9+\i,-0.8) -- +(0,1.6);
		
		\foreach \i in {-0.05,0.05} \draw (1.9+\i,-0.8) -- +(0,1.6);
		\foreach \i in {-0.05,0.05} \draw (3.1+\i,-0.8) -- +(0,1.6);
		
		\foreach \i in {1,2}{ \node[anchor=south] at ({-0.4+0.8*\i},0.8) {\footnotesize $k_{\i}$};}
		\node[anchor=south] at (3.95,0.8) {\footnotesize $k_{L}$};
		
		\node at (-0.5,-0.35) {\footnotesize $N_0$};
		\node at (0.65,-0.35) {\footnotesize $N_1$};
		\node at (1.4,-0.37) {\footnotesize $N_2$};
		\node at (4.15,-0.37) {\footnotesize $N_{L}$};
		\node at (4.8,-0.37) {\footnotesize $N_{L+1}$};  
		\draw[->] (4.3,-1)--node[above]{$x_3$}(4.7,-1);
	\end{tikzpicture}
	\caption{Brane construction of an interface CFT, which can be truncated to a BCFT or 3d SCFT by turning off one or both of $N_0$ and $N_{L+1}$. Vertical ellipses: NS5-branes. Horizontal lines: D3-branes extending in the $x_3$ direction, with $N_t$ denoting the number between adjacent NS5-branes. Vertical lines: D5-branes intersecting D3-branes, with $k_t$ denoting the number. }\label{fig: brane-set-up}
\end{figure}
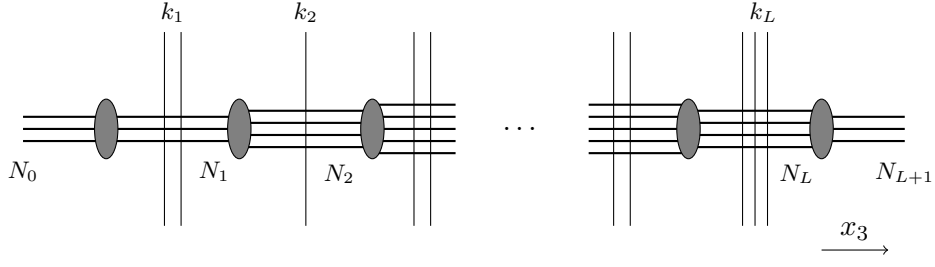

For the discussion of $\frac{1}{2}$-BPS boundaries, we take $\mathbb R^{1,3}$ restricted to the half space $x^3\ge 0$ and decompose the four-dimensional spacetime indices accordingly into $\mu=(i,3)$, with $i=0,1,2$. Half-BPS interface theories are covered by this discussion via the folding trick \cite{Gaiotto:2008sa}. The preserved superconformal symmetry is 3d $\man=4$, i.e.\ $OSp(4|4)$ with bosonic subalgebra $SO(2,3) \times SO(3)\times SO(3)$, where the R symmetry is $SO(3)\times SO(3)$.
The 4d vector multiplet $(A_\mu, \Phi_I,\Psi)$ 
splits into a 3d vector multiplet  and a 3d  hypermultiplet
\bea
\text{Vector: }& A_{i},Y_i,\Psi_+~, \\
\text{Hyper: }& A_{3},X_i,\Psi_-~,
\eea
where the 4d scalars are split into triplets $\Phi_I=(X_i,Y_i)$ and the gaugino into $\Psi_{\pm} = \frac{1}{2}\left(1 \pm \Gamma_{3456}\right)\Psi$.
There is a rich class of boundary conditions preserving 3d $\mathcal N=4$ supersymmetry. These include as relatively simple examples the following:
\begin{itemize}
	\item \textbf{Dirichlet / Nahm pole (D5 type)}: The vector multiplet satisfies the Dirichlet boundary condition
	\bea
	F_{\mu\nu}| = Y_i | = \Psi_{+}| = 0~,
	\eea
	while the hypermultiplet satisfies a generalized Neumann boundary condition
	\bea
	D_3 X_i - \frac{i}{2}\epsilon_{ijk}[X_j, X_k] | =0~,
	\eea
	known as Nahm pole boundary condition since $X_i$ satisfies the singular Nahm equation near the boundary. The solutions take the form
	\bea
	X_i=\frac{t^i}{x_3}+\text{regular}~.
	\eea
	This boundary condition is realized by D3-branes ending on D5-branes.
	The standard Dirichlet boundary condition corresponds to the trivial representation $t^i=0$, where each D3-brane ends on one D5-brane.
	$N$ D3-branes ending on a single D5-brane corresponds to $t_i$ forming the $N$-dimensional irreducible representation of $SU(2)$.
	In general, for groups of $K_i$ D3-branes with each group ending on one D5-brane, $t^i$ corresponds to a reducible representation of $SU(2)$ with $K_i \times K_i$ blocks.
	Despite the fact that no explicit new 3d fields are introduced, these boundary conditions already introduce 3d boundary degrees of freedom \cite{Gaiotto:2008sa,Gaiotto:2008ak}.
	
	\item \textbf{Neumann (NS5 type)}: The vector multiplet satisfies the Neumann boundary condition
	\bea
	F_{3\mu}| = D_{3}Y^{i}| = 0~,
	\eea
	while the hypermultiplet satisfies the Dirichlet boundary condition
	\bea
	X^{i}| =\Psi_{-}|=0~.
	\eea
	This is realized by D3-branes ending on a single NS5-brane. More general NS5-type boundary conditions involving multiple NS5-branes introduce explicit 3d fields.
\end{itemize}
The general $\frac{1}{2}$-BPS boundary/gluing conditions involve 3d theories with $\mathcal N=4$ superconformal symmetry.
They are realized by D3-branes with 10d orientation (0123) ending on or intersecting groups of D5 (012456) and NS5-branes (012789), as shown in figure \ref{fig: brane-set-up}.\footnote{More general boundary conditions can be realized with $(p,q)$ 5-branes and/or $SL(2,\mathbb{Z})$ rotations.}
The 3d defect theories live on the $(012)$ directions shared by the D3/D5/NS5 branes.
D3-branes stretched between NS5-branes represent 3d gauge nodes \cite{Hanany:1996ie}, and the resulting quivers are classified into good, bad and ugly according to the behavior in the IR limit \cite{Gaiotto:2008ak}.
We focus on good theories, where the 3d theory flows to an IR SCFT without decoupled sectors.
This IR limit corresponds to taking all 5-branes coincident.
Semi-infinite D3-branes stretching beyond the 5-branes represent 4d SYM on half spaces.
Depending on the presence of semi-infinite D3-branes, one obtains three classes of theories: 3d SCFTs, 4d BCFTs, and 4d interface CFTs (ICFTs). These correspond, respectively, to the cases in which both, one, or neither of the two numbers $N_0$ and $N_{L+1}$ vanish.
General ICFTs can be represented by quiver diagrams from which the matter content can be read off directly \cite{Giveon:1998sr}.
For the ICFTs represented by the brane configuration in figure~\ref{fig: brane-set-up}, the quiver takes the form~\footnote{Strictly speaking, the D3/D5 interface theory which is subject of section \ref{sec: D3/D5} is not covered by this form of the quiver diagram, though the S-dual D3/NS5 interface is described by $\widehat{U(N)} - U(N) -...-U(N)- \widehat{U(N)}$.}
\begin{align}\label{eq:3d4d-quiver}
	\begin{array}{ccccccccccc}
		\widehat{U(N_0)} & - & U(N_1) & - & U(N_2) & - & \cdots & - & U(N_L) & - & \widehat{U(N_{L+1})} \\
		&   & |      &   & |      &   &        &   & |       &   & \\
		&   & [k_1]  &   & [k_2]  &   & \cdots &   & [k_L]   &   &
	\end{array}
\end{align}
\begin{itemize}
	\item $\widehat{U(\cdot)}$ nodes represent 4d SYM on half spaces, engineered by semi-infinite D3-branes;
	
	\item $U(\cdot)$ nodes represent 3d gauge nodes, engineered by finite D3-brane stacks stretching between neighboring NS5-branes. Each 3d node contributes a 3d $\man=4$ vector multiplet, arising from open strings ending on the associated D3-brane segments;
	
	\item the gauge nodes are connected by bifundamental hypermultiplets arising from open strings stretched between neighboring D3-brane stacks, shown as dashes;
	\item $[k_t]$ denote flavor symmetries, with the vertical dashes representing hypermultiplets in the fundamental representation of the connected gauge node, arising from open strings stretching between D3-brane segments and D5-branes. 
\end{itemize}
An alternative brane realization can be obtained by moving all D5-branes to one side and all NS5-branes to the other side by Hanany-Witten transitions, as explained e.g.\ in \cite{Hanany:1996ie,Assel:2011xz}, but we will not use this form.

In the remainder of this section we apply the formalism set up in section \ref{sec: local} to a sample of boundary, defect and interface theories based on $\mathcal N=4$ SYM ambient theories and preserving 3d $\mathcal N=4$ superconformal symmetry.

\subsection{\texorpdfstring{$\man=4$}{N=4} Janus}\label{sec:N4Janus}

The first example is the pure Janus interface, which is the simplest four-dimensional ICFT obtained by allowing the Yang-Mills coupling of $\mathcal N=4$ SYM to jump across a codimension-one plane, without changing the rank of the ambient gauge groups or introducing additional dynamical degrees of freedom localized on the interface. 

The initial constructions were non-supersymmetric, while preserving the full $SO(6)$ R-symmetry of $\mathcal N=4$ SYM, with the holographic solution constructed in \cite{Bak:2003jk} and identified as the dual of an interface CFT in \cite{Clark:2004sb}. That some supersymmetry could be preserved at the expense of breaking R-symmetry was noted in \cite{Clark:2004sb,Clark:2005te}, and the options were classified in \cite{DHoker:2006qeo}. We focus here on the maximally supersymmetric option with 3d $\mathcal N=4$ supersymmetry, for which holographic duals were constructed in \cite{DHoker:2006vfr,DHoker:2007zhm}.

The one-point function of the Lagrangian operator for this theory was studied recently in \cite{Karch:2026ymg} in the limits of weak jump using conformal perturbation theory and at strong coupling using holography. Remarkably, the results agreed, suggesting that the one-point function could in fact be exact. Here we show that this is indeed the case.

For the pure $\mathcal{N}=4$ Janus interface, the matrix model on $S^4$ with the interface along an equatorial $S^3$ differs from the matrix model for standard $\mathcal N=4$ SYM only in the weight of the Gaussian, encoding the coupling dependence \cite{Goto:2020per,He:2025due},
\bea \label{eq: Janus-matrix}
\mathcal Z&=\int \prod_{i=1}^N da_i \prod_{i<j}(a_i-a_j)^2\exp\left\lbrace -\frac{8\pi^2 N}{\overline{\lambda}}\sum_{i=1}^Na_i^2\right\rbrace,
\eea
where 
\bea
\overline{\lambda}\equiv \frac{2\lambda_L\lambda_R}{\lambda_L+\lambda_R} ~,
\eea
and $\lambda_L\equiv g_L^2N$ and $\lambda_R\equiv  g_R^2N$ are the 't Hooft couplings on the two sides of the interface. 
This model arises by gluing two 4d $\mathcal N=4$ SYM partition functions on $HS^4$ using the formulas of \cite{Dedushenko:2018tgx} with trivial kernel.
The interface-localized terms required to preserve supersymmetry do not contribute on the localization locus.

Computing the free energy is identical to standard $\mathcal N=4$ SYM, see e.g.\ \cite{Raamsdonk:2020tin}. The result is
\begin{align} \label{eq: Janus-free-en}
\mathcal{F}\left(\lambda_L,\lambda_R\right)
= -\frac{N^2}{2}\log \overline{\lambda} + K(N) ~,
\end{align}
where $K(N)$ is independent of the couplings. The result for standard $\mathcal N=4$ SYM is recovered by setting $\lambda_L=\lambda_R$.
Using this in the definition of the interface free energy in \eqref{eq: def-free-en-gen} leads to
\begin{align}
\mathcal{F}_{\mathrm{def}}
&=\mathcal{F}\left(\lambda_L,\lambda_R\right) -\frac{1}{2}\mathcal{F}\left(\lambda_L,\lambda_L\right)-\frac{1}{2}\mathcal{F}\left(\lambda_R,\lambda_R\right)
\nonumber\\
&=
 -\frac{N^2}{2}\,\log\frac{2\sqrt{\lambda_L\lambda_R}}{\lambda_L+\lambda_R} =-\frac{N^2}{4}\,\log\left(1-\gamma^2\right) ~,
 &
 \gamma \equiv
 \frac{\lambda_L-\lambda_R}{\lambda_L+\lambda_R} ~,
\end{align}
up to coupling-independent constants, and with $\gamma$ as jump parameter.
This agrees with the result obtained in \cite{Goto:2020per} by supergravity methods.

The one-point function of the Lagrangian density can be extracted from derivatives of the interface free energy. Since we have independent gauge couplings on the two sides of the interface, the relevant relation is the half-space formula in \eqref{eq: dF-to-L}.
When differentiating with respect to the left/right coupling we keep $\lambda_{L/R}$ fixed. We find
\begin{align}
\mathcal{C}_{\mathcal O_{\mathcal L}}^{L/R}=-\frac{3}{4\,\pi^2}\lambda_{L/R}\,\partial_{\lambda_{L/R} }\mathcal F_{\mathrm {def}} = \mp\frac{3N^2}{16\pi^2}\,\gamma ~.
\label{eq: aLJfin}
\end{align}
As a consistency check, we can compare \eqref{eq: aLJfin} to the one-point function of the chiral primary operator $\mathcal{O}_2$. From \cite{He:2025due} we have
\bea
\mathcal{C}_{\mathcal{O}_2}^{L/R} = \frac{N}{2\sqrt{2}} \left(2\,{}_2F_1\left(-1,1;2;\frac{\overline{\lambda}}{\lambda_{L/R}}  \right)  -1\right) = \pm\frac{N}{2\sqrt{2}}\gamma ~,
\label{aOJfin}
\eea
where the sign depends on which side the one-point function is computed on. Combining the two results \eqref{eq: aLJfin} and \eqref{aOJfin}, we immediately recover the relation \eqref{eq:COL-CO2}.

The localization result (\ref{eq: aLJfin}) is exact in $N$ and in the couplings $\lambda_L$ and $\lambda_R$.
In the appropriate limits it agrees with the results from holography and conformal perturbation theory obtained in \cite{Karch:2026ymg}, while extending to intermediate $N$ and $\lambda_{L/R}$. Our result in (\ref{eq: aLJfin}) therefore establishes the exactness property conjectured in \cite{Karch:2026ymg}.

\subsection{D3/D5 defects}

\label{sec: D3/D5}

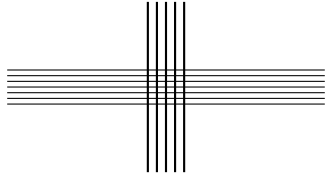
\begin{figure}
\centering
		\begin{tikzpicture}[yscale=0.75]
			\pgfmathsetmacro{\s}{0.3}
			\pgfmathsetmacro{\t}{0.1}
			
			\foreach \i in {-2,...,2} \draw[thick] (0.4*\i*\s,-1.5) -- +(0,3);
			\foreach \j in {-3,-2,-1,0,1,2,3} \draw (-7*\s,\j*\t) -- (7*\s,\j*\t);
		\end{tikzpicture}
	\caption{Defect in $\mathcal N=4$ SYM realized by D5-branes intersecting  D3-branes.  \label{fig: D3D5-brane}}
\end{figure}

As a second example we consider the half-BPS defect in $\mathcal N=4$ SYM with gauge group $U(N)$ engineered by intersecting $N_{\rm D5}$ D5-branes with $N$ D3-branes \cite{Karch:2000gx,DeWolfe:2001pq,Aharony:2003qf}, illustrated in fig.~\ref{fig: D3D5-brane}. The D5-branes introduce $N_{\rm D5}$ three-dimensional hypermultiplets localized on the defect, transforming in the fundamental representation of $U(N)$. The corresponding matrix model was proposed in \cite{Robinson:2017sup} and later derived in \cite{Wang:2020seq}. For $N_{\rm D5}=1$, the one-point functions of chiral primary operators $\mathcal O_J$ were computed at arbitrary value of the coupling in \cite{Komatsu:2020sup,Beccaria:2022bjo}. More recently, these one-point functions were studied in the regime with a large number of D5-branes, $N_{\rm D5}\sim N$, in the supergravity limit in \cite{He:2025due}. The matrix model reads
\bea 
\mathcal Z&=\int \prod_{i=1}^N da_i \prod_{i<j}(a_i-a_j)^2\exp\left\lbrace -\frac{8\pi^2 N}{\lambda}\sum_{i=1}^Na_i^2\right\rbrace \displaystyle \prod_{j=1}^N \left( 2\cosh (\pi\,a_j) \right)^{-N_{D5}} ~.
\label{mmd3d5}
\eea
We will derive the one-point function of the Lagrangian density in two distinct regimes. The first, $N_{\rm D5}=1$, allows us to keep the full dependence on the 't Hooft coupling and hence to probe finite-coupling effects beyond the supergravity approximation. The second, $N_{\rm D5}\sim N$, is the regime of fully backreacted 5-branes.

\subsubsection*{Single D5-brane}

For a single D5 the defect free energy and one-point functions of chiral primary operators, $\mathcal C_{\mathcal{O}_J}$, were studied in the planar limit in \cite{Komatsu:2020sup}. These results were subsequently extended to all orders in the large-$N$ expansion in \cite{Beccaria:2022bjo}, by exploiting the integrable Volterra hierarchy underlying the matrix model \cite{Alvarez-Gaume:1990asn}. The resulting differential relations provide a systematic way to generate the $1/N$-corrections to both the defect free energy and the one-point functions $\mathcal C_{\mathcal{O}_J}$. Here we focus on the planar limit and derive the one-point function of the Lagrangian density for arbitrary coupling. The non-planar corrections can be obtained by applying our formula \eqref{eq: dF-to-L} to the large-$N$ expansion of the defect free energy.
 
The planar defect free energy can be written as \cite{Komatsu:2020sup,Beccaria:2022bjo}
\bea
\mathcal{F}_{\rm def} = \frac{4\,N}{\pi}\int_{0}^{\frac{\pi}{2}} \!d\theta\,\sin^2\!\theta \,\log \left[ 2\cosh\!\left(\frac{\sqrt{\lambda}}{2}\cos\theta  \right) \right] ~.
\eea
Using \eqref{eq: dF-to-L}, this immediately implies
\bea
\mathcal{C}_{\mathcal O_{\mathcal{L}}} = -\frac{3\,N\sqrt{\lambda}}{8\,\pi^3}\int_{0}^{\frac{\pi}{2}} \!d\theta\,\sin^2\theta \,\cos\theta\,\tanh\!\left(\frac{\sqrt{\lambda}}{2}\cos\theta  \right) ~.
\label{eq: L1d3d5}
\eea
This expression is exact in the 't Hooft coupling. In contrast to the Janus interface, the weak- and strong-coupling expansions are genuinely different:
\begin{align}
\mathcal{C}_{\mathcal O_{\mathcal{L}}} \underset{\lambda\ll 1}{=} & \,\, \frac{N}{\pi^2}\!\left( -\frac{3\,\lambda}{256} +\frac{\lambda^2}{2048} -\frac{\lambda^3}{32768}+\frac{17\,\lambda^4}{7864320} + O\!\left(\lambda^5  \right) \right) , \\[0.5em]
\mathcal{C}_{\mathcal O_{\mathcal{L}}} \underset{\lambda\gg 1}{=} & \,\, N\!\left( -\frac{\sqrt{\lambda}}{8\,\pi^3} +\frac{1}{16\,\pi\sqrt{\lambda}} -\frac{7\,\pi}{320\lambda^{3/2}}-\frac{31\,\pi^3}{2688\lambda^{5/2}} + O\!\left(\lambda^{-7/2}  \right) \right) . \label{eq: strongexp}
\end{align}
We can also verify the relation between the exact expressions for $\mathcal{C}_{\mathcal O_{\mathcal{L}}}$ and $\mathcal{C}_{\mathcal{O}_2}$. In \cite{Komatsu:2020sup} the  one-point function of the chiral primary operator $\mathcal{O}_2$ was found to be
\bea
\mathcal{C}_{\mathcal{O}_2} = \frac{\sqrt{2}}{2\,\pi}\int_{-\frac{\pi}{2}}^{\frac{\pi}{2}} \!d\theta\,\cos(2\theta) \,\log \left[ 2\cosh\!\left(\frac{\sqrt{\lambda}}{2}\cos\theta  \right) \right] ~.
\eea
Integrating by parts, we end up with the following expression 
\bea
\mathcal{C}_{\mathcal{O}_2} = \frac{\sqrt{2\,\lambda}}{2\,\pi}\int_{0}^{\frac{\pi}{2}} \!d\theta\,\sin^2\theta \,\cos\theta\,\tanh\!\left(\frac{\sqrt{\lambda}}{2}\cos\theta  \right)  ~.
\eea
Comparing this form with \eqref{eq: L1d3d5}, we find precise agreement with the identity in \eqref{eq:COL-CO2}.

\subsubsection*{Large-$N_{D5}$}
When the number of D5-branes is large, $N_{\rm D5} \sim N/\sqrt{\lambda}$, the D5-brane contribution in the matrix model \eqref{mmd3d5} backreacts and changes the saddle point substantially. The saddle-point equation was solved in \cite{He:2025due}, where the corresponding defect free energy was also obtained in the supergravity limit as
\bea
\mathcal{F}_{\rm def}=\frac{N^2}{12} \left( 9-d^4-8d^2-12\log d \right) ~,
\eea
with
\bea
d\equiv\sqrt{1+\mathfrak{N}_{\mathrm{D5}}^2}-\mathfrak{N}_{\mathrm{D5}}  \quad \text{and} \quad \mathfrak{N}_{\mathrm{D5}}\equiv\frac{N_{\mathrm{D5}}\sqrt{\lambda}}{4\pi\,N}~.
\label{eq: ddef}
\eea
We can thus evaluate the one-point function of the Lagrangian density using the formula \eqref{eq: dF-to-L}
\bea
\mathcal{C}_{\mathcal O_{\mathcal{L}}} =  \frac{N^2}{16\pi^2}\left(d^2-1\right)\left(d^2+3\right) ~. 
\label{eq: Ld3d5}
\eea
The expansion of this expression to linear order in $N_{\rm D5}$ reproduces the leading strong-coupling result for the single-D5 case quoted in \eqref{eq: strongexp}. By comparing the expression \eqref{eq: Ld3d5} with the corresponding coefficient $\mathcal{C}_{\mathcal{O}_2}$ computed in \cite[eq.(63)]{He:2025due}, it is straightforward to verify that the relation \eqref{eq:COL-CO2} is satisfied.

\subsection{D3/NS5 BCFTs}\label{sec:D3NS5BCFT}

We now turn to a BCFT for which the free energy has been computed exactly. It is engineered by $N\equiv N_5K$ D3-branes ending on a group of $N_5$ NS5-branes, with $K$ D3-brane terminating on each NS5.
In field theory terms, the BCFT arises as IR fixed point of 4d $\mathcal N=4$ SYM on a half space, coupled to a 3d $\man=4$ quiver gauge theory. In the language of the quiver diagram (\ref{eq:3d4d-quiver}) it takes the form
\begin{align}\label{eq:D3NS5-BCFT-quiver}
	&U(K)-U(2K)-U(3K)-\ldots - U((N_5-1)K)-\widehat{U(N_5K)}~.
\end{align}
The matrix model is 
\begin{align}\label{eq:D3NS5-BCFT-matrix-model}
	\mathcal Z=\,& \frac{1}{N_{1} ! \ldots N_{L} !} \int\left(\prod_{t=1}^{L} \prod_{i=1}^{N_t} d a_{t,i}\right) 
	e^{-\frac{4 \pi^2}{g_{\mathrm{YM}}^2} \sum_{i=1}^{N_L} a_{L, i}^2} \prod_{i<j}^{N_L}\left(a_{L, i}-a_{L, j}\right) 2\sinh\left(\pi\left(a_{L, i}-a_{L, j}\right) \right)
	\nonumber\\
	& \prod_{t=1}^{L-1} \prod_{i<j}^{N_t} 4\sinh^2\left(\pi\left(a_{t,i}-a_{t,j}\right) \right)
	\prod_{t=1}^{L-1} \prod_{i=1}^{N_t} \prod_{j=1}^{N_{t+1}} \frac{1}{2\cosh\left(\pi\left(a_{t,i}-a_{t+1,j}\right)\right)}~,
\end{align}
where $L=N_5$ and $N_t=t K$.
The boundary free energy has been obtained using localization with full dependence on $N$ and $\lambda$ in \cite{Raamsdonk:2020tin}. The result is 
\begin{align}
\mathcal F_{\mathrm{def}}
=&
-\left(
\frac{N(N-N_5)}{2}
+\frac{N}{4}
\right)\ln(2\pi)
-\frac{N^2}{4}\ln\left(\frac{\lambda}{4\pi^2 N}\right)
-\frac{\lambda}{48}\left(\frac{N^2}{N_5^2}-1\right)
\nonumber\\
&-N_5^2 \ln G_2\left(\frac{N}{N_5}+1\right)
+\frac{1}{2}\ln G_2(N+1)~,
\end{align}
where $G_2 (N+1)=\prod_{k=1}^{N-1} k!$ is the Barnes G-function.
With the relation (\ref{eq: dF-to-L}), we immediately obtain the one-point function of the Lagrangian with full dependence on $N$ and $\lambda$ as 
\begin{equation}\label{eq: one-point-BCFT-gen}
\mathcal{C}_{\mathcal O_{\mathcal{L}}}=\dfrac{3N^2}{16\pi^2}
+\dfrac{\lambda}{64 \pi^2 }\left(\frac{N^2}{N_5^2}-1\right).
\end{equation}
In the `t Hooft limit, $N\rightarrow\infty$ with fixed $\lambda\gg 1$ and fixed large $N_5$ so that $N_5^2\sim\lambda$, this reduces to
\begin{equation}\label{eq: one-point-BCFT-gra}
\mathcal{C}_{\mathcal O_{\mathcal{L}}}=\dfrac{N^2}{16\pi^2}
	\left(3+\dfrac{\lambda}{4N_5^2}\right).
\end{equation}
Comparing \eqref{eq: one-point-BCFT-gra} with \cite[eq.(86)]{He:2025due}, we can verify that the relation \eqref{eq:COL-CO2} between $\expval{\mo_2}$ and $\expval{\mathcal O_{\mathcal L}}$ holds in this theory as well.

\subsection{B/I/dCFTs with balanced 3d matter}

For boundary, defect and interface CFTs with 3d matter described by balanced quiver diagrams, i.e.\ where the number of flavors for each 3d gauge node equals twice the number of colors, we can give a general closed form for the one-point functions of the Lagrangian in the holographic limit. The results in this subsection are restricted to this holographic limit, where the matrix models are dominated by saddle points and the one-point functions of $\mathcal O_2$ were determined in \cite{He:2025due}.
Special cases include the D3/D5/NS5 BCFTs, which have found applications in the study of black holes and double holography \cite{Uhlemann:2021nhu,Demulder:2022aij,Karch:2022rvr,Billiato:2025jkr}, and the S-dual of the D3/D5 defects discussed in section \ref{sec: D3/D5}.

The general form of the quiver diagram is in (\ref{eq:3d4d-quiver}), and the assumption of balanced nodes means $k_t+N_{t-1}+N_{t+1}=2N_t$ for all $t$ corresponding to 3d nodes, i.e.\ $t=1,..,L$.
The NS5-brane number is $N_5=L+1$.
If $N_0$ and $N_{L+1}$ are both non-zero, we have an interface or defect CFT, and if one of them vanishes we have a BCFT. Taking $N_0=N_{L+1}=0$ would produce a genuine 3d SCFT, but this case is not of interest here. The matrix model reads 
\begin{align}\label{eq: bcft-matrix}
\mathcal Z=\,& \frac{1}{N_{0} ! \ldots N_{L+1} !} \int\left(\prod_{t=0}^{L+1} \prod_{i=1}^{N_t} d a_{t,i}\right) 
e^{-\frac{4 \pi^2}{g_L^2} \sum_{i=1}^{N_0} a_{0, i}^2}
e^{-\frac{4 \pi^2}{g_R^2} \sum_{i=1}^{N_{L+1}} a_{L+1, i}^2}
\nonumber\\
&\prod_{i<j}^{N_0}\left(a_{0, i}-a_{0, j}\right) 2\sinh\left(\pi\left(a_{0, i}-a_{0, j}\right) \right)
\prod_{i<j}^{N_{L+1}}\left(a_{L+1, i}-a_{L+1, j}\right) 2\sinh\left(\pi\left(a_{L+1, i}-a_{L+1, j}\right) \right)
\nonumber\\
& \prod_{t=1}^{L}\prod_{i<j}^{N_t} 4\sinh^2\left(\pi\left(a_{t,i}-a_{t,j}\right) \right)
\prod_{t=0}^{L}\prod_{i=1}^{N_t} \prod_{j=1}^{N_{t+1}} \frac{1}{2\cosh\left(\pi\left(a_{t,i}-a_{t+1,j}\right)\right)}
\prod_{t=1}^{L}\prod_{i=1}^{N_t} \frac{1}{2^{k_t}\cosh^{k_t}(\pi a_{t,i})}.
\end{align}
The one-point function for the chiral primary operators $\mathcal O_J$ were derived in \cite[eq.\ (128)]{He:2025due} from a saddle point analysis of this matrix model, and we will use the relation between $\langle \mathcal O_2\rangle$ and the one-point function of $\mathcal O_{\mathcal L}$ in (\ref{eq:COL-CO2}) to derive the one-point function of the Lagrangian. The one-point functions for $\mathcal O_J$ in \cite[eq.\ (128)]{He:2025due} simplify for $J=2$ to
\begin{align}\label{eq:one-point-NS5-int-gen}
	\langle{\mathcal O}_2\rangle
	&=\frac{1}{-2\sqrt{2}}\biggr[N_{\rm D3}+
	 N_5\Re \Res_{u=-1}\Bigg[
	\nonumber\\
	&
	 \frac{dv}{du} \left( K_0\frac{1-u}{1+u}-K_1\frac{1+u}{1-u}-\frac{i}{\pi}\sum_{p=1}^P N_{\rm D5}^{(p)}\ln\left(\frac{\frac{1-u}{1+u}-i e^{\delta_p}}{\frac{1-u}{1+u}+ie^{\delta_p}}\right) \right)T_2\left(\frac{N_5(v-i\pi)}{\sqrt{\lambda}}\right)\Bigg]\Bigg],
\end{align}
with $T_2(x)=-1+2x^2$ and $v=\frac{\pi K_2}{N_5} \frac{1-u}{1+u}+\frac{\pi K_3}{N_5} \frac{1+u}{1-u}-\ln u$.
The residue can be evaluated explicitly for $J=2$ and the one-point function of $\mathcal O_{\mathcal L}$ follows from (\ref{eq:COL-CO2})
\bea
\mac_{\mathcal O_{\mathcal L}}^L
=\frac{1}{64\pi^2}\Bigg\{
&K_0^2
\left[
3\left(\pi K_3+2N_5\right)^2
+4\pi K_2 N_5
\right]
\\
&-
3K_2^2\Bigg(
\pi K_1
+2\sum_{p=1}^{P}
N_{\mathrm{D5}}^{(p)}e^{\delta_p}
\Bigg)^2
-4\pi K_0K_2^2
\sum_{p=1}^{P}
N_{\mathrm{D5}}^{(p)}e^{3\delta_p}
\Bigg\}~.
\eea
The quantities $K_{0,1,2,3}$, $N_{\rm D5}^{(p)}$ and $\delta_p$ encode the field theory parameters $(g_{L/R}, N_{0,L+1}, k_t)$, where each $N_{\rm D5}^{(p)}$ gives a flavor group and $\delta_p$ encodes which gauge node it is attached to. The conversion is given by \cite[(115)]{He:2025due} and the two-sided version of \cite[(48)]{He:2024djr},
\bea
i\pi\left(\frac{t_p}{N_5}-1\right)=\frac{i\pi}{N_5}\left(K_2 e^{\delta_p}-K_3 e^{-\delta_p}\right)+\log\left(\frac{i e^{\delta_p}+1}{i e^{\delta_p}-1}\right)~.
\eea
where $t_p$ identifies the gauge node which the $p^{\rm th}$ flavor group is attached to.

We discuss two special cases as examples. The first is the D3/D5/NS5 BCFT with one flavor group used in \cite{Uhlemann:2021nhu,Karch:2022rvr}. It is specified by $K_1 = K_3 =\delta=0$.
Upon introducing the combinations
\begin{align}
\mathfrak{N}_5 &= \frac{N_5}{\sqrt{\lambda}}~,
&
\mathfrak{N}_{\mathrm{D}5} &= \frac{N_{\mathrm{D}5}\sqrt{\lambda}}{4\pi N_{\mathrm{D}3}}~,
&
\mathfrak{N}_{\pm} &=  \mathfrak{N}_{\mathrm{D}5} \pm \mathfrak{N}_5 ~,
\end{align}
the remaining parameters in the one-point function can be expressed as
\begin{align}
K_0 &= \frac{N_{\mathrm{D}3}}{\sqrt{\lambda}\,\mathfrak{N}_{+}}~,
&
K_2 &= \frac{\sqrt{\lambda}}{4\pi \mathfrak{N}_{+}}~.
\end{align}
Then the one-point function becomes
\bea
\mac_{\mathcal O_{\mathcal L}}^L
&=-\frac{N_{\rm D3}^2\mathfrak N_-}{64\pi^2}
\frac{12\mathfrak N_+^2+1}{\mathfrak N_+^3}~.
\eea
The second example is the D3/NS5 defect, where $K_0 = K_1$, $K_2 = K_3$ and $P=0$.
Then $K_{0,2}$ can be inverted into field theory parameters as
\begin{align}
K_0&=\frac{2N}{\sqrt{\lambda}}d~,
&
K_2&=\frac{\sqrt{\lambda}}{2\pi}d~,
\end{align}
with 
\bea
d\equiv\sqrt{1+\mathfrak{N}_5^2}-\mathfrak{N}_5
\quad \text{and} \quad
\mathfrak{N}_5 \equiv \frac{N_5}{\sqrt{\lambda}}~.
\label{eq: ddefNS5}
\eea
The one-point function becomes
\bea
\mac_{\mathcal O_{\mathcal L}}
&=\frac{N^2}{16\pi^2}(1-d^2)(d^2+3)~.
\eea
We can compare this to the result for the backreacted D5 defect in (\ref{eq: Ld3d5}): noting that the definition of $d$ there is different from the definition of $d$ here, and that the two are exchanged under S-duality, we find that the one-point functions are mapped into each other.

\subsection{Non-renormalization properties}\label{sec:non-ren}

For the $\frac{1}{2}$-BPS Janus interface it was shown in \cite{Karch:2026ymg} that the one-point function $\langle \mathcal O_{\mathcal L}\rangle$ computed to first order in conformal perturbation theory agrees with the holographic result, which is valid in the planar limit with large but independent couplings on the two sides of the interface. 
Our results in section \ref{sec:N4Janus} show that, as one may expect, the one-point function (\ref{eq: aLJfin}) is exact, i.e.\ valid for arbitrary $N$ and $\lambda_{L/R}$.

One may wonder if this exactness should not extend to general defect and interface theories with 3d $\mathcal N=4$ superconformal symmetry, i.e.\ whether there is a general non-renormalization theorem protecting the one-point function. This is not the case: our results, e.g.\ for the D3/D5 defect theory in section \ref{sec: D3/D5} or for the D3/NS5 BCFT in section \ref{sec:D3NS5BCFT}, clearly show non-trivial coupling dependence. The exactness of the leading order calculation in conformal perturbation theory is a special feature of the Janus interface.

Some perspective on what makes the Janus interface special can be gained from the matrix model description, as we elaborate now.
The matrix model $\eqref{eq: Janus-matrix}$ has the same general form as that of standard $\man=4$ SYM \eqref{eq: SYM-matrix}: the one-loop determinants cancel between the $\mathcal N=2$ vector multiplet and hypermultiplet, and there are no instanton contributions. The entire coupling-dependence of the partition function can be factored out by a rescaling of the eigenvalues; setting
\bea\label{eq:scaling-a}
a_i = \overline{\lambda}^{\frac{1}{2}}\, a'_i
\eea
isolates the coupling dependence in an overall factor and leads to
\bea
\mathcal Z (\overline \lambda)=\overline{\lambda}^{\frac{N^2}{2}}Z (1)~.
\eea
The free energy \eqref{eq: Janus-free-en} then follows immediately. In particular, it has the same form at weak and strong coupling and for large and small $\lambda_L-\lambda_R$. Together with the relation (\ref{eq: dF-to-L}) this explains the agreement between conformal perturbation theory and holographic results for the one-point function of $\mathcal O_{\mathcal L}$ in this theory.

For general Gaiotto-Witten theories with 3d matter the matrix models are clearly more involved. They in particular contain non-trivial one-loop determinants $Z_{\rm 1-loop}\!\left(\{a_{t,i}\}\right)$ in \eqref{eq:BCFT-matrix-model-gen}, which arise from the 3d fields. These schematically take the form 
\bea
Z_{\rm 1-loop}\!\left(\{a_{t,i}\}\right)\sim \sinh^{\alpha} (a_{t,i})\cosh^{\beta} (a_{t,i})~,
\eea
where the integer powers $\alpha$ and $\beta$ are determined by the matter content of the theory (compare (\ref{eq: bcft-matrix})). While the rescaling in (\ref{eq:scaling-a}) still suffices to trivialize the coupling dependence of the Gaussian 4d $\mathcal N=4$ SYM contribution, it now generates non-trivial coupling dependence in the 3d one-loop determinants. As a result, the defect free energy and one-point function $\langle \mathcal O_{\mathcal L}\rangle$ have non-trivial coupling dependence and are not fully captured by either conformal perturbation theory or holography. While the above argument strictly speaking applies to theories with explicit 3d matter, it extends by S-duality to D3/D5 interfaces producing Nahm pole boundary conditions without explicit 3d matter.

\section{\texorpdfstring{$\langle \mathcal O_{\mathcal L}\rangle$}{} in Gaiotto-Witten theories holographically}\label{sec: holo}

In this section we obtain the one-point functions of the Lagrangian density in Gaiotto-Witten theories from their holographic duals. This gives direct access to the local operator, as the dual to the dilaton, and complements the supersymmetric localization approach which proceeds through integrated correlators. It will provide useful consistency checks and further non-trivial tests of the dualities: while many consistency checks of the relevant dualities have been performed, e.g.\ at the level of partition functions, extended operators, and through connections to matrix models, we extend this here to local operators.

\subsection{\texorpdfstring{$\ads_4 \times S^2 \times S^2 \times \Sigma$}{} solutions}
\label{sec: holo duals}
The holographic duals of the Gaiotto-Witten theories are half-BPS solutions of Type IIB supergravity, constructed in \cite{DHoker:2007zhm,DHoker:2007hhe,Aharony:2011yc,Assel:2011xz} (see also \cite{Coccia:2021lpp} for a compact review).
The geometry is a warped product of $\ads_4\times S^2\times S^2$ over a 2d Riemann surface $\Sigma$, with non-vanishing dilaton, NS-NS 2-form and R-R 2- and 4-forms.
The Einstein-frame metric and dilaton are
\begin{align}\label{eq:ds2-IIB}
	ds^2 = f_4^2\, ds^2_{\ads_4}+f_1^2\, ds^2_{S_1^2}+f_2^2\, ds^2_{S_2^2}
	+4\rho^2 |dz|^2~,
\end{align}
where the warp factors $f_4,f_1,f_2$ and $\rho$ depend on the coordinates of the
2d Riemann surface $\Sigma$.
The solutions are parametrized by two harmonic functions $h_1$ and $h_2$ defined on $\Sigma$.
From these one constructs composite quantities
\begin{align}
	W &= \partial\bar{\partial}(h_1 h_2)~, 
	& N_i &= 2h_1h_2 |\partial h_i|^2 - h_i^2 W~,
\end{align}
and the metric functions are then given by
\begin{align}
	f_4^8 &= 16\frac{N_1 N_2}{W^2}~, 
	& f_1^8 &= 16 h_1^8 \frac{N_2 W^2}{N_1^3}~, 
	& f_2^8 &= 16 h_2^8 \frac{N_1 W^2}{N_2^3}~, 
	&
	\rho^8 &= \frac{N_1 N_2 W^2}{h_1^4 h_2^4}~.
\end{align}
Among the 10d fields, the one relevant for our purposes is the dilaton.
We adopt the convention of
\cite{DHoker:2007zhm,DHoker:2007hhe}, where the axion-dilaton field is written
as $\tau=\chi+i e^{-2\phi}$ with
\begin{align}\label{eq:fieldstrengths-IIB}
	e^{4\phi} &= \frac{N_2}{N_1}~,
\end{align}
and vanishing axion.
We parametrize the 2d Riemann surface $\Sigma$ as an infinite strip with complex coordinate $(z,\bar z)$
\begin{align}
	\Sigma&=\left\lbrace z\in \mathds{C}\, \vert \, 0\leq\Im(z)\leq \frac{\pi}{2}\right\rbrace~.
\end{align}
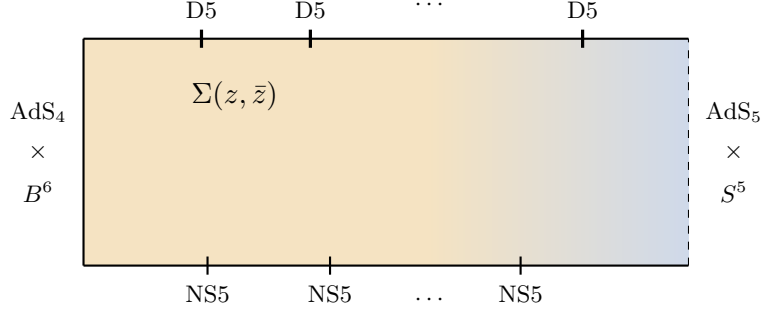
\begin{figure}
	\centering
	\begin{tikzpicture}[xscale=2,yscale=1.5]
		\shade [right color=3dcolor!100,left color=3dcolor!100] (-0.3,0)  rectangle (0.3,-2);
		
		\shade [ left color=3dcolor! 100, right color=4dcolor! 100] (0.3-0.01,0)  rectangle (2,-2);
		\shade [ right color=3dcolor! 100, left color=3dcolor! 100] (-0.3+0.01,0)  rectangle (-2,-2);
		
		\draw[thick] (-2,0) -- (2,0);
		\draw[thick] (-2,-2) -- (2,-2);
		\draw[dashed] (2,-2) -- +(0,2);
		\draw[thick] (-2,-2) -- +(0,2);
		
		\node at (-1,-0.5) {$\Sigma(z,\bar z)$};
		\node at (-2.3,-0.65) {\footnotesize $\rm AdS_4$};
		\node at (-2.3,-1) {\footnotesize $\times$};
		\node at (-2.3,-1.35) {\footnotesize $B^6$};			
		\node at (2.3,-0.65) {\footnotesize $\rm AdS_5$};
		\node at (2.3,-1) {\footnotesize $\times$};
		\node at (2.3,-1.35) {\footnotesize $S^5$};

		\foreach \i in {-1.1,-0.5,1}{
			\draw[very thick] (1.2*\i+0.1,-0.08) -- (1.2*\i+0.1,0.08) node [anchor=south] {\footnotesize D5};}
		\node at (0.3,0.3) {\footnotesize $\cdots$};
		\foreach \i in {-1.2,-0.3,1.1}{
			\draw[thick] (0.9*\i-0.1,-1.92) -- (0.9*\i-0.1,-2.08) node [anchor=north] {\footnotesize NS5};
		}			
		\node at (0.3,-2.3) {\footnotesize $\cdots$};
	\end{tikzpicture}
	\caption{The 2d strip $\Sigma$ in Type IIB solutions dual to 4d BCFTs. Groups of D5- and NS5-brane sources are on the top and bottom boundary components, respectively. The geometry caps off smoothly on the left, while it blows up to $\ads_5 \times S^5$ on the right. }\label{fig: strip-bcft}
\end{figure}
The harmonic functions on this strip are given by
\begin{align}\label{eq:h1h2-gen}
	h_1&=-\frac{i \pi \alpha^\prime}{4} (K_0 \, e^z-K_1 \, e^{-z})-\frac{\alpha^\prime}{4} \sum_{a=1}^AN_{\rm D5}^{(a)}\ln\tanh\left(\frac{i\pi}{4}-\frac{z-\delta_a}{2}\right)+\rm{c.c.}~,
	\nonumber\\
	h_2&=\frac{\pi \alpha^\prime}{4} (K_2 \,  e^z+K_3 \,e^{-z})-\frac{\alpha^\prime}{4}\sum_{b=1}^B N_5^{(b)}\ln\tanh\left(\frac{z-\delta_b}{2}\right)+\rm{c.c.}~.
\end{align}
These solutions describe D3-branes suspended between, ending on, or intersecting D5- and NS5-branes. Specifically, $A$ groups of D5-branes with $N_{\rm D5}^{(a)}$ D5-branes in the $a^{\rm th}$ group located at $z=\delta_a + i\frac{\pi}{2}$, and $B$ groups of NS5-branes with $N_5^{(b)}$ NS5-branes in the $b^{\rm th}$ group located at $z=\delta_b$ \cite{Assel:2011xz}.
The field theory is a 4d interface CFT when $K_{0,1,2,3}\neq 0$, a 4d BCFT when either $K_{0,2}\neq 0$ or $K_{1,3}\neq0$, and a 3d SCFT when $K_{0,1,2,3}=0$.
We present the strip $\Sigma$ for $K_{0,2}\neq 0$ and $K_{1,3}=0$ dual to a 4d BCFT in figure \ref{fig: strip-bcft}.
Near the boundaries $\Re{z}\to\pm\infty$,
there is an asymptotic $\ads_5\times S^5$ region whenever the dual CFT includes a corresponding 4d SYM on the half space; otherwise it closes off smoothly with the vertical boundary of the strip becoming a regular interior point of the 10d geometry.

\subsection{One-point functions}

We now compute the one-point function of the operator dual to the zero-mode of the dilaton on $S^5$. This operator is $\mathcal O_{\mathcal L}$. The one-point function is extracted from the asymptotic expansion of the dilaton at the conformal boundaries emerging at $\Re(z)=\pm \infty$.
In the asymptotic region at $\Re(z)=\pm \infty$ we introduce real coordinates $(r,y)$ with
\begin{equation}
	z=\mp \log r +i y~,
\end{equation}
so that the asymptotic region corresponds to $r\rightarrow 0$. When the exponential terms in (\ref{eq:h1h2-gen}) do not vanish in the asymptotic region, the 10d metric becomes 
\bea\label{eq:asympt-ds2}
ds^2=& L_{\pm}^2\left(\frac{1}{r^2}\left(dr^2+\frac{ds^2_{\ads_4}}{b_{\pm}^2}\right)+ds^2_{S^5}\right)+\cdots~
\eea 
where $L_\pm$ and $b_{\pm}$ are constants and the ellipsis denotes terms which are subleading at small~$r$. The conformal boundary in this parametrization is (two copies of) $\ads_4$. In order to extract the one-point functions on flat space, we define a new radial coordinate combining $r$ with the $\ads_4$ radial coordinate, namely we fix $\ads_4$ metric and radial coordinate $f$ as
\begin{align}
	ds^2_{\ads_4}&=\frac{-dt^2+dx_1^2+dx_2^2+dx_3^2}{x_3^2}~,
	&
	f&=\,b_{\pm}\,x_3\,r~.
\end{align}
Then the conformal boundary near $f\to 0$ is a half plane.
In order to read off the one-point function of the Lagrangian from the expansion of $\phi$, we in principle need to ensure that the metric is in Fefferman-Graham gauge to the required order in $r$.
For the dilaton case at hand, fixing the leading behavior in (\ref{eq:asympt-ds2}) turns out to suffice, since the first $r$-dependent term will be the desired one-point function.

With these preparations the one-point function $\expval{\mo_{\ml}}$ is given by the subleading term in the near boundary expansion of the dilaton,
\bea \label{eq: holo-dic}
\phi=\phi^{(0)}+\frac{\pi^2}{N^2}\expval{\mo_{\ml}} f^4+\cdots~,
\eea
where $\phi^{(0)}$ is identified with the $\mathcal N=4$ SYM gauge coupling through $e^{2\phi^{(0)}}=g^2/(4\pi)$. This takes into account the dilaton normalization used in \cite{DHoker:2007zhm,DHoker:2007hhe} where $\tau = \chi + ie^{-2\phi}$ (and differs from \eqref{eq: FGexp}).
The coefficient $\frac{\pi^2}{N^2}$ in the relation of the $f^4$ term to the one-point function arises from the volume of $S^5$ \cite{Balasubramanian:1998de}.
In general, holographic renormalization \cite{Bianchi:2001kw,Skenderis:2002wp} is needed to extract the one-point functions from the near boundary expansion of the dual bulk fields, with counterterms built from local functions of the sources.
Because the dilaton is dual to a marginal operator, all counterterms involve derivatives of the sources. They consequently vanish for a constant source and are not required for this computation.

For the quantitative analysis we specialize to configurations with a single group of NS5-branes at $z=0$ and a single group of D5-branes at $z=i\frac{\pi}{2}$.
The harmonic functions \eqref{eq:h1h2-gen} then simplify to
\begin{align}\label{eq:h1h2-our}
	h_1&=-\frac{i \pi \alpha^\prime}{4} (K_0 \, e^z-K_1 \, e^{-z})-\frac{\alpha^\prime}{4} N_{\rm D5}\ln\tanh\left(\frac{i\pi}{4}-\frac{z}{2}\right)+\rm{c.c.}~,
	\nonumber\\
	h_2&=\frac{\pi \alpha^\prime}{4} (K_2 \,  e^z+K_3 \,e^{-z})-\frac{\alpha^\prime}{4} N_5\ln\tanh\left(\frac{z}{2}\right)+\rm{c.c.}~.
\end{align}
This will allow for a number of field theory comparisons, and the generalization to generic $h_{1/2}$ as in (\ref{eq:h1h2-gen}) is straightforward.
The asymptotic value of the dilaton sets the gauge coupling and the number of semi-infinite D3-branes can be extracted from the radii of the emerging $\ads_5\times S^5$ region through the relations
\begin{align}
	\lim_{\Re(z)\rightarrow \pm\infty}e^{2\phi}&=\frac{g_{L/R}^2}{4\pi}~,
	 & L_{\pm}^4 &=4\pi \ap'^2 N_{L/R}~,
\end{align} 
where $g_{L/R}$ is the SYM coupling of the left/right ambient theory.
With the harmonic functions in (\ref{eq:h1h2-our}) we find for the $\Re(z)=+\infty$ asymptotic region 
\begin{align} \label{eq: field-to-gravity}
e^{2\phi}\big\vert_{z=+\infty}&=\frac{K_2}{K_0} ,
& 
N_{L} &=\frac{\pi}{2}(K_1 K_2 + K_3 K_0)+K_0 \ns+K_2 \nd~.
\end{align}
The constant relating $f$ and $r$ is $ b_+^2 = \frac{4N_{\rm L}}{\pi K_0K_2}$.
The expansion of the dilaton takes the form
\begin{align}
\phi\big\vert_{\Re(z)\gg 1}
&=
\frac{1}{2}\log\frac{K_2}{K_0}
+\frac{\pi^2}{N_L^2}\,
\frac{\mac_{\mathcal O_{\mathcal L}}^L}{x_3^4}\,f^4+\cdots~,
\end{align}
where
\begin{align} \label{eq: one-point-gravity-gen}
\mac_{\mathcal O_{\mathcal L}}^L
&=
\frac{3K_0^2\left(\pi K_3+2\ns\right)^2
	-3K_2^2\left(\pi K_1+2\nd\right)^2
	+4\pi K_0K_2\left(K_0\ns-K_2\nd\right)}{64\pi^2}
\,.
\end{align}
Comparing to \eqref{eq: holo-dic}, the coefficient $\mac_{{\mathcal O_{\mathcal{L}}}}^L$ is the desired one-point function coefficient.
The results for the $\Re(z)\rightarrow -\infty$ asymptotic region are obtained by exchanging $K_{0,2}\leftrightarrow K_{1,3}$.

To compare with field theory results, one needs to convert the supergravity parameters $K_{0,1,2,3}$ into field theory parameters $(N_{L/R}\,,g^2_{L/R})$ by inverting \eqref{eq: field-to-gravity} and the analog for the second asymptotic region.
We now specialize the general result for the holographic one-point functions, in order to compare to the theories discussed in section \ref{sec: SYM}:
\begin{itemize}
	\item The Janus interface is recovered for $N_{\rm D5}=N_5=0$. This leads to 
\begin{align}
	\mac_{\mathcal O_{\mathcal{L}}}^L
	&=
	\frac{3(K_0^2K_3^2
		-K_2^2K_1^2)}{64}~.
\end{align}
Expressing the supergravity parameters in terms of field theory quantities using \eqref{eq: field-to-gravity} and $ \gamma =
 \frac{\lambda_L-\lambda_R}{\lambda_L+\lambda_R}$ reproduces \eqref{eq: aLJfin}.

\item The D3/D5 defect theory is recovered by setting $N_5=0$ and $K_{1,3}=K_{0,2}$.
Then \eqref{eq: one-point-gravity-gen} becomes
\bea \label{eq: one-point-DCFT-hol}
\mac_{\mathcal O_{\mathcal{L}}}&=-\frac{K_2^2 N_{\rm D5}\left(3N_{\rm D5}+4\pi K_0\right)}{16\pi^2}~.
\eea
We then express $K_0$, $K_2$ and $N_{\rm D5}$ in terms of the field theory parameters using \eqref{eq: field-to-gravity} and \eqref{eq: ddef},
\bea \label{eq: one-point-DCFT-hol-subs}
K_2=\frac{\sqrt{\lambda}d}{2 \pi}~,\qquad K_0=\frac{2 N d}{\sqrt{\lambda}}~,\qquad N_{\rm D5} = -\frac{2 \pi N \left(d^2-1\right)}{\sqrt{\lambda}d}~.
\eea
Substituting \eqref{eq: one-point-DCFT-hol-subs} into \eqref{eq: one-point-DCFT-hol}, we find exact agreement with the field theory result in the supergravity limit \eqref{eq: Ld3d5}.

\item
The D3/NS5 BCFT is recovered for $K_1=K_3=N_{\rm D5}=0$. Replacing further $K_0, K_2$ with field theory parameters using \eqref{eq: field-to-gravity} leads to
\bea \label{eq: eq: one-point-BCFT-hol}
\mac_{\mathcal O_{\mathcal{L}}}&=\frac{K_0^2 N_5(3N_5+\pi K_2)}{16\pi^2} =\dfrac{N^2}{16\pi^2}
\left(3+\dfrac{\lambda}{4N_5^2}\right)~.
\eea
This exactly agrees with the field theory results in the supergravity limit in (\ref{eq: one-point-BCFT-gra}).

\end{itemize}
Naturally, the holographic result for the Lagrangian one-point function can be compared to the field theory results for $\langle\mathcal O_2\rangle$ obtained in \cite{He:2025due} by using the relation \eqref{eq:COL-CO2}.
The supergravity solution specified by \eqref{eq:h1h2-our} is dual to the D3/D5/NS5 interface represented by the quiver \eqref{eq:3d4d-quiver} with one group of fundamental hypermultiplets. 
For this theory $\expval{ \mathcal O_2}$ can be obtained in \cite[eq.(128)]{He:2025due} with $P=1$ and $\delta_p=0$. As expected, this is consistent with the relation \eqref{eq:COL-CO2}, providing a further consistency check.

\section{Janus interfaces with reduced supersymmetry}\label{sec:Janus-less-susy}

In this section we discuss Janus interfaces in $\mathcal N=4$ SYM with reduced supersymmetry. We label them by the amount of 3d defect superconformal symmetry, so that $\mathcal N=4,2,1$ correspond to interfaces in 4d $\mathcal N=4$ SYM preserving, respectively, $\frac{1}{2}$, $\frac{1}{4}$, $\frac{1}{8}$ of the supersymmetry. We will not reduce the supersymmetry of the ambient theory.
The interfaces in 4d $\mathcal N=4$ SYM were classified in \cite{DHoker:2006qeo}.
The cases we focus on are the $\mathcal N=2$ interface with $SU(2)$ flavor symmetry, the $\mathcal N=1$ interface with $SU(3)$ flavor symmetry, for which holographic duals were provided in \cite{Bobev:2020fon}, and the $\mathcal N=0$ interface of \cite{Bak:2003jk,Clark:2004sb}.
We first discuss the matrix model for $\mathcal N=2$ Janus, before discussing $\mathcal N=2,1,0$ holographically.

\subsection{One-point functions in localization}\label{sec:N2Janus-loc}

In this section we return to 4d $\mathcal N=2$ SYM coupled to hypermultiplets and, importantly, with position-dependent complex coupling $\tau$, as realized in \cite{Goto:2020per}. This construction already played a role in section~\ref{sec:O2-OL}. Generic smooth profiles for $\tau$ can be seen as regulator for Janus interfaces, which emerge when the coupling profile approaches a step function.

Here we discuss a simple consequence of the localization computations in \cite{Goto:2020per}. The authors realize 4d $\mathcal N=2$ SYM with position-dependent coupling $\tau$ on $S^4$ with an off-shell supersymmetry, so that the usual localization arguments can be applied. This leads to the same localization locus and one-loop determinants as in standard $\mathcal N=2$ SYM, since these are determined solely by the localizing term $\delta V$: a real scalar in the vector multiplet takes a constant value on $S^4$, parametrizing a matrix integral. Remarkably, the modifications to the matrix integral due to the position-dependent coupling can be captured by simply continuing $(\tau,\bar\tau)$ to $(\tau_+,\bar\tau_-)$, where $\tau_\pm$ are the values of $\tau$ at the poles of $S^4$. The precise profile of the coupling on $S^4$ is not relevant. The matrix model becomes
\begin{equation}
	\mathcal Z_\text{4d $\mathcal N=2$ SYM}(\tau_+,\bar\tau_-)=\int [da] e^{-I_{\rm cl}(\tau_+,\bar\tau_-)}Z_{\rm 1-loop}(a) Z_{\rm inst}(a,e^{2\pi i \tau_+})Z_{\rm inst}(a,e^{-2\pi i\bar\tau_-})\,,
\end{equation}
where the classical action evaluated on the localization locus is
\begin{equation}
	I_{\rm cl}(\tau_+,\bar\tau_-)=-i\pi r^2(\tau_+-\bar\tau_-)\tr a^2~.
\end{equation}
Conceptually, computations can now be performed with smooth coupling profiles, taking the sharp interface limit at the end. Though the profile does not enter the matrix model, this is useful for the discussion of interface terms which enhance or break supersymmetry.

To realize interfaces in $\mathcal N=4$ SYM, a 4d $\mathcal N=2$ adjoint hypermultiplet has to be added. As reviewed in section~\ref{sec:O2-OL} this is straightforward since hypermultiplets do not directly couple to the multiplet hosting the position-dependent coupling. In the matrix model the hypermultiplet contributions cancel the one-loop determinant of the vector multiplet and trivialize the instanton partition functions, which only depend on the couplings at the poles, following the arguments in \cite{Pestun:2007rz} for standard $\mathcal N=4$ SYM. The hypermultiplet does not contribute to the classical action on the localization locus and the matrix model becomes
\begin{equation}
	\mathcal Z_\text{4d $\mathcal N=4$ SYM}(\tau_+,\bar\tau_-)=\int [da] e^{-I_{\rm cl}(\tau_+,\bar\tau_-)} \,.
\end{equation}
The resulting field theory preserves 3d $\mathcal N=2$ supersymmetry, with the scalars grouped into one complex scalar in the 4d $\mathcal N=2$ vector multiplet and two complex scalars in the 4d $\mathcal N=2$ hypermultiplet. It has a $U(1)$ R-symmetry and an $SU(2)$ flavor symmetry acting on the hypermultiplet scalars. This is case (II) of the $\mathcal N=2$ interfaces in \cite{DHoker:2006qeo}. 

In summary, the matrix model for an interface in 4d $\mathcal N=4$ SYM preserving 3d $\mathcal N=2$ defect superconformal symmetry with an $SU(2)$ flavor symmetry becomes Gaussian and identical to the matrix model for $\mathcal N=4$ Janus discussed in section~\ref{sec:N4Janus}. The one-point functions of the chiral primary operator $\langle\mathcal O_2\rangle$ and ambient Lagrangian $\langle\mathcal O_{\mathcal L}\rangle$ should therefore agree with those of the maximally supersymmetric Janus interface.

\medskip

We now discuss Janus interfaces in $\mathcal N=4$ SYM with 3d $\mathcal N=4$ or 3d $\mathcal N=1$ supersymmetry. These can be realized by adding terms proportional to derivatives of the coupling, which do not change the ambient theory. One may hope to incorporate these terms in all cases as simple deformations of the matrix model action, but this turns out to be subtle.

Keeping the terms proportional to coupling variations well-defined in the sharp-interface limit is non-trivial, and can be accomplished by coupling-dependent field rescalings \cite{DHoker:2006qeo}. In the sharp-interface limit the fields are rescaled by step functions, which dictates the gluing conditions for finite-action field configurations: in the variables where the interface terms are regular, the notion of finite-action field configurations reduces to the usual notion of smoothness. With $\phi^i$ denoting the six real $\mathcal N=4$ SYM scalars in a normalization where the coupling appears as overall $1/g^2$ in the action, the rescalings are
\begin{align}
	\mathcal N_{3d}&=4 : & \phi^i&=g^2\tilde \phi^i\,,\!\!\qquad i=2,4,6~,
	\nonumber\\
	\mathcal N_{3d}&=2 : &  \phi^{i}&=g\tilde\phi^{i}\,,\qquad i=3,4,5,6~,\qquad \phi^2=g^2\tilde\phi^2~,
	\nonumber\\
	\mathcal N_{3d}&=1 : &  \phi^i&=g\tilde\phi^i\,,\qquad i=1,..,6
\end{align}
For the $\mathcal N_{3d}=2$ case, $\phi^{3,4,5,6}$ are the scalars in the 4d $\mathcal N=2$ hypermultiplet, and the rescalings convert the kinetic term to canonical normalization. Among the remaining 4d $\mathcal N=2$ vector multiplet scalars $\phi^{1,2}$, $\phi^1$ is continuous across the interface and can be identified with the real scalar whose constant value parametrizes the localization locus.

For interfaces with the maximal 3d $\mathcal N=4$ defect superconformal symmetry, the scalar $\phi^1$ remains unrescaled and continuous across the interface. The interface terms can be extracted from \cite{DHoker:2006qeo} by comparing the terms for $\mathcal N=4$ and $\mathcal N=2$ interfaces in (5.6) and (5.11), respectively. There are no new terms involving $\phi^1$, so that the additional interface terms do not contribute on the localization locus. The matrix model remains unchanged, in line with the discussion in section~\ref{sec:N4Janus} and consistent with the results there.

For $\mathcal N=1$ interfaces, on the other hand, all scalars are rescaled and jump across the interface. The localization locus thus does not correspond to a finite-action field configuration with the interface terms for 3d $\mathcal N=1$ supersymmetry and the localization results do not extend straightforwardly. We will therefore only discuss this case holographically.

\subsection{\texorpdfstring{$\mathcal N=2,1,0$}{} Janus holographically}

We now discuss the one-point functions in $\mathcal N=2,1,0$ Janus interfaces holographically, for $\mathcal N=0$ using the solutions of \cite{Bak:2003jk}, and for $\mathcal N=1$ Janus with $SU(3)$ flavor symmetry and $\mathcal N=2$ Janus with $SU(2)$ flavor symmetry using the solutions of \cite{Bobev:2020fon}.

The supersymmetric solutions are formulated in consistent truncations to 5d gauged supergravities with effective Newton constant $G_5=\frac{\pi L^3}{2 N^2}$. The geometry takes the form 
\begin{equation} \label{eq: metric-AdS5-AdS4}
ds^2_5 =dr^2+e^{2 A(r)} ds^2_{\ads_4}~,
\end{equation}
with a warp factor $A(r)$ depending on the solution and determined by BPS or field equations in conjunction with a set of scalar fields. Type IIB string theory solutions are obtained by uplifting these 5d solutions to 10d.

The Lagrangian operator $\mathcal O_{\mathcal L}$ is dual to the 10d dilaton $\Phi$. Crucially, $\Phi$ is in general not identified with any particular field in the 5d supergravity description, but rather emerges upon uplifting to 10d as a non-trivial combination of the 5d fields. The asymptotic value of the 10d dilaton at $r\rightarrow \pm\infty$ sets the gauge couplings on the two half spaces via
\begin{align}
	\lim_{r\rightarrow\pm\infty}\mathrm{e}^{\Phi}=\frac{g_{L/R}^2}{4\pi}~,
	\label{eq: dictionary}
\end{align}
and we will denote the coupling jump throughout as
\begin{equation}
	\gamma\equiv\frac{g_L^2-g_R^2}{g_L^2+g_R^2}=\tanh\!\left(\frac{\Phi_+-\Phi_-}{2}\right)~.
	\label{eq: gamma}
\end{equation}
The expectation value of $\mathcal O_{\mathcal L}$ is extracted from the near-boundary expansion of $\Phi$ in Fefferman-Graham gauge. To describe the field theory on flat space (rather than two copies of $AdS_4$) the radial coordinate should be identified as
\begin{equation}
	f=z\mathrm e^{-A}~,
	\label{eq: FG-def}
\end{equation}
where $z$ is the Poincar\'e radial coordinate of the AdS$_4$ slice, $ds^2=z^{-2}(dz^2+ds^2_{\RR^{3}})$.
The one-point functions are then obtained as subleading term in the near boundary expansion
\begin{align}
	\mathrm e^{\Phi}=\mathrm e^{\Phi_\pm}\left[1+\frac{2\pi^2}{N^2}\big\langle \mathcal O_{\mathcal L} \big\rangle f^4+\cdots\right]~.
	\label{eq: FGexp}
\end{align}   

In the following we holographically compute $\langle \mathcal O_{\mathcal L}\rangle$ for the $\mathcal N=2,1,0$ Janus interfaces. We also compute the defect free energies and will verify the relation \eqref{eq: dF-to-L} between the coupling derivative of the free energy and $\langle \mathcal O_{\mathcal L}\rangle$ for all interfaces.

\subsubsection{\texorpdfstring{$\mathcal{N}=2$}{} Janus}

The $\mathcal N=2$ Janus solution is constructed in a 5d truncation involving the metric and 6 scalar fields denoted $(\varphi,\alpha,\chi,\lambda,c,\omega)$ in \cite{Bobev:2020fon}.
We first review the solution and then discuss the one-point function.
The solution is expressed in terms of an alternative radial variable $X(r)$, which is determined from the first-order equation
\begin{align}\label{eq:X-eq}
	\left(\frac{dX}{dr}\right)^2+V_{\rm eff} =0~,
\end{align}
describing a particle in an effective potential $V_{\rm eff}$.
Combining equations (3.14) and (3.16) in \cite{Bobev:2020fon}, while setting $L=1$, yields the expression for the warp factor in terms of the radial coordinate $X$ as
\begin{align}
	\mathrm e^{-2A}=\frac{2}{\sqrt{\mathcal I}}\frac{\sqrt{X}}{\left( 1-X \right)^{1/6}}  ~.
	\label{eq: warp}
\end{align}
where $0<\mathcal{I}< 1$ is an integral of motion, which controls the Janus deformation parameter $\gamma$.
The 5d metric \eqref{eq: metric-AdS5-AdS4} becomes
\begin{align}\label{eq:5dmetric-N2Janus}
	ds_5^2 = \frac{dX^2}{-V_{\rm eff}(X)} + \frac{\sqrt{\mathcal I}\left( 1-X \right)^{1/6}}{2\sqrt{X}} ds^2_{\rm AdS_4} ~.
\end{align}
The effective potential is
\begin{align}
	V_{\rm eff}(X) =-\frac{16(1-X)^{1/3}X^2}{\sqrt{\mathcal I}}\left(\sqrt{\mathcal I}-2\sqrt{X(1-X)}\right)~.
	\label{eq: Veff}
\end{align}
The turning point for $X(r)$ is where the potential vanishes and given by
\begin{align}
	X_{\rm tp} =\frac12\left(1-\sqrt{1-\mathcal I}\right)~.
	\label{eq: tp}
\end{align}
Both $AdS_5$ regions emerge for $X\rightarrow 0$.
The 5d dilaton $\varphi$ takes the form
\begin{equation}
	\cosh 2\varphi=\cosh 2F+\frac12 \mathrm{e}^{-2F}\mathcal J^2 ~,
	\label{eq: varphi-F}
\end{equation}
where $\mathcal J$ is a second integral of motion, and
\begin{align}
	F(X)&=F_0\pm\int_{X_{\rm tp}}^X \!\dd x\,\frac{(3-2x)x}{\sqrt{\mathcal I}(1-x)^{4/3}+2(1-x)^{5/6}x^{3/2}}\frac{1}{\sqrt{-V_{\rm eff}(x)}}~.
	\label{eq: F-int}
\end{align}
The two signs are exchanged at the turning point. As explained in section 2 of \cite{Bobev:2020fon}, a constant global $SL(2,\mathbb R)$ transformation can be used to set
\begin{equation}
	(\varphi, c,\omega)\longrightarrow(F,0,0)\quad \Longrightarrow\quad \mathcal J=0~.
	\label{eq: frame}
\end{equation}
We work in this duality frame from now on.
With $\varphi_\pm$ denoting the two asymptotic values of the 5d dilaton, we define $\eta$ as half their difference as
\begin{align}
	\eta\equiv{}& \frac{\left|\varphi_+-\varphi_-\right|}{2} 
	=\int_0^{X_{\rm tp}}\frac{\dd x}{\sqrt{-V_{\rm eff}(x)}}\,\frac{(3-2x)x}{\sqrt{\mathcal I}(1-x)^{4/3}+2(1-x)^{5/6}x^{3/2}}~.
	\label{eq: eta}
\end{align}
The integration endpoints follow from the fact that both asymptotic AdS$_5$ regions correspond to $X=0$, while the two branches meet at $X=X_{\rm tp}$.
This is a fully explicit one-dimensional integral which can easily be evaluated numerically. We find
\begin{align}
	\eta = \frac{1}{2}\mathrm{arctanh}\left(\sqrt{\mai}\right)~.
	\label{eq: res-eta}
\end{align}
The remaining scalar fields $\alpha$, $\lambda$, $\chi$ are determined from \cite[(3.11),(3.12),(3.16)]{Bobev:2020fon} as
\begin{align}\label{eq:alpha-chi-lambda}
	\mathrm e^{6\alpha}&=\sqrt{1-X}~,
	&
	\sinh^2(2\chi) &=\frac{2X^{3/2}}{\sqrt{\mathcal I}\sqrt{1-X}}~,
	&
	\mathrm e^{6\alpha}\cosh2\lambda\cosh2\chi=1~.
\end{align}
The uplifted 10d axion and dilaton are given in \cite[(3.34)]{Bobev:2020fon} as
\begin{align}\label{eq:uplift-tau}
	\mathrm e^\Phi&=\frac{1}{2\sqrt{CK_1K_2}}\Bigl[2CK_3+\mathrm e^{-6\alpha}\sin^2\theta\bigl(d_--C^2d_++S^2\cos2\phi\bigr)\Bigr]~,
	\nonumber\\
	C_0&=\frac{2C\cos^2\theta\sinh2\varphi-\frac12\mathrm e^{-6\alpha}\sin^2\theta\,\partial_\lambda(d_--C^2d_+)}{2CK_3+\mathrm e^{-6\alpha}\sin^2\theta\bigl(d_--C^2d_++S^2\cos2\phi\bigr)}~,
\end{align}
where $\theta$ and $\phi$ are coordinates on the internal $S^5$ and
\begin{align}
	C&=\cosh2\chi~, & b_\pm&=\cosh2\lambda\pm\sin2\phi\,\sinh2\lambda~,\label{eq:CS}
	\nonumber\\
	S&=\sinh2\chi~, & d_\pm&=\cosh(2\lambda\pm2\varphi)-\sin2\phi\,\sinh(2\lambda\pm2\varphi)~,
\end{align}
and
\begin{align}
	K_1&=\cos^2\theta+C\mathrm e^{-6\alpha}b_-\sin^2\theta~, &
	K_2&=C\cos^2\theta+\mathrm e^{-6\alpha}b_-\sin^2\theta~, \nonumber\\
	K_3&=\cos^2\theta\cosh2\varphi+C\mathrm e^{-6\alpha}d_+\sin^2\theta~. \label{eq:K}
\end{align}

\paragraph{One-point function} To extract the one-point function we need the asymptotic expansion of the 10d dilaton and axion. 
For the one-point function $\big\langle \mathcal O_{\mathcal{L}}\big\rangle$, we need the  coefficient of the fourth power of the Fefferman-Graham coordinate $f$, defined in \eqref{eq: FG-def}. From the warp factor \eqref{eq: warp}
\begin{align}
	f^4=z^4\mathrm e^{-4A}=\frac{4z^4 X}{\mathcal I(1-X)^{1/3}} \qquad \Longrightarrow\qquad X=\frac{\mathcal I}{4z^4}f^4+O(f^8)~.
	\label{eq: X-FG}
\end{align}
Therefore, we must keep all the $X$-dependence up to $O(X)$.
In the frame defined in \eqref{eq: frame}, the 5d scalars relevant to the axio-dilaton uplift are $(\varphi,\alpha,\chi,\lambda)$. Since $\varphi$ approaches the finite values $\varphi_\pm$, we keep it unexpanded for the uplift. 
The scaling of the remaining scalar fields with $X$ near the asymptotic region $X\rightarrow 0$ can be read off from (\ref{eq:alpha-chi-lambda}) as
\begin{align}\label{eq:alpha-chi-lambda-scale}
	\alpha&=O(X)~,
	&
	\chi&=O(X^{3/4})~,
	&
	\lambda=O(X^{1/2})~.
\end{align}

Near either asymptotic AdS$_5$ region, the Fefferman-Graham coordinate is related to $X$ by \eqref{eq: X-FG}. Determining the $f^4$ coefficient of the 10d axio-dilaton using the uplift (\ref{eq:uplift-tau}) requires retaining every contribution through order $X$. Terms starting at $O(X^{3/2})=O(f^6)$ may be neglected only after verifying that the uplift formulae contain no inverse powers of $X$ that could enhance them.  We therefore first establish the power counting of all quantities entering \eqref{eq:uplift-tau}.
Given the scaling of the scalar fields in \eqref{eq:alpha-chi-lambda-scale}, we readily have
\begin{align}
	C&=1+O(X^{3/2})~, &  S^2&=O(X^{3/2})~, 
&
	\mathrm e^{-6\alpha}&=1+\frac{X}{2}+O(X^2)~.
	\label{eq:CSexpalpha-scale}
\end{align}
For the remaining quantities appearing in \eqref{eq:uplift-tau} we get:
\begin{align}
	b_\pm&=1+O(X^{1/2})~,&
	d_\pm&=\cosh2\varphi\mp\sin2\phi\,\sinh2\varphi+O(X^{1/2})~,\nonumber\\
	K_1&=1+O(X^{1/2})~, \nonumber\\
	K_2&=1+O(X^{1/2})~, &
	K_3&=\cosh2\varphi-\sin2\phi\sin^2\theta\sinh2\varphi+O(X^{1/2})~. \label{eq: K3-scale}
\end{align}
With these expressions both uplift formulae are non-singular as $X\to0$: their denominators remain non-zero, and the coefficients multiplying $C-1$ or $S^2$ remain finite.  Therefore the dependence of the complete axio-dilaton on $C-1$ and $S^2$ starts at $O(X^{3/2})=O(f^6)$ and cannot affect the  $f^4$ term. Through the required order we may consequently replace
\begin{equation}
	C\longrightarrow1~, \qquad S^2\longrightarrow0~,
	\label{eq: replacement}
\end{equation}
while keeping the complete dependence on $\alpha$, $\lambda$ and $\varphi$.  In particular, $b_-$, $d_\pm$ and $K_i$ are not replaced by their leading boundary values but are kept in their exact form in the following.
After the replacements \eqref{eq: replacement}, $K_1$ and $K_2$ coincide, and we have 
\begin{align}
	K_1=K_2=\cos^2\theta+\mathrm e^{-6\alpha}b_-\sin^2\theta~.
	\label{eq: K}
\end{align}
We can then exploit the exact identities
\begin{align}
	d_-+d_+=2b_-\cosh2\varphi~, \qquad \partial_\lambda(d_--d_+)=-4b_-\sinh2\varphi ~,
	\label{eq: identities}
\end{align}
to remove all remaining dependence on $\alpha$, $\lambda$ and the internal coordinates. We obtain the 10d dilaton and axion as
\begin{align}
	\tau=C_0+i\mathrm e^{-\Phi}=\tanh2\varphi+i\sech2\varphi+O(X^{3/2})~.
	\label{eq: final-uplift}
\end{align}

In order to describe interfaces with vanishing $\theta$-angle, where only the gauge coupling jumps, we apply an $SL(2,R)$ transformation to set the asymptotic value of the axion to zero. This is accomplished by
\begin{align}
	\tau'=\frac{\tau-1}{\tau+1}~,
\end{align}
which leads to
\begin{align}
	\tau'=i \mathrm e^{-2\varphi} +O(X^{3/2})~.
	\label{eq: final-tau}
\end{align}
The final relation between the 5d dilaton and the physical 10d dilaton becomes
\begin{align}
	\Phi=2\varphi +O(X^{3/2})~.
	\label{eq: final-dilaton}
\end{align}
Using \eqref{eq: gamma}, \eqref{eq: eta}, \eqref{eq: res-eta} and \eqref{eq: final-dilaton} we find the coupling jump $\gamma$ as
\begin{align}\label{eq:gamma-N2}
	\gamma=\tanh\!\left(\frac{\Phi_+-\Phi_-}{2}\right)=\tanh\left(\mathrm{arctanh}\left( \sqrt{\mathcal I} \right)\right)=\sqrt{\mathcal I}~.
\end{align}
We note that the factor two in the uplift (\ref{eq: final-dilaton}) is crucial for this simple relation to emerge.
We now extract the $f^4$ term.
The expansion of the 5d dilaton near the asymptotic regions can be determined from the integrand of \eqref{eq: F-int} as
\begin{align}
	\varphi(r\to\pm\infty)=\varphi_\pm\mp\frac{3}{4\sqrt{\mathcal I}}X+O(X^{3/2})~.
	\label{eq: varphi-X-exp}
\end{align}
From \eqref{eq: X-FG} and \eqref{eq: final-dilaton} we immediately find
\begin{align}
	\mathrm e^{\Phi}
	=\mathrm e^{\Phi_\pm}
	\left[1\mp\frac{3\sqrt{\mathcal I}}{8z^4}f^4+O(f^6)\right]~.
	\label{eq: expPhi-exp}
\end{align}
Comparing \eqref{eq: FGexp} with \eqref{eq: expPhi-exp}, and using $z=|x^3|$, we finally obtain
\begin{align}
	\langle {\mathcal O_{\mathcal L}}\rangle=-\frac{3N^2\sqrt{\mathcal I}}{16\pi^2}\frac{\sgn(x^3)}{|x^3|^4} = -\frac{3N^2\gamma}{16\pi^2}\frac{\sgn(x^3)}{|x^3|^4} ~.
	\label{eq: 1ptL-N2}
\end{align}
This is identical to the result for the $\mathcal{N}=4$ Janus interface in (\ref{eq: aLJfin}), in line with the localization discussion in section~\ref{sec:N2Janus-loc}.

\subsubsection{\texorpdfstring{$\mathcal{N}=1$}{} Janus}

The $\mathcal N=1$ Janus solution is constructed in a truncation of the theory hosting the $\mathcal N=2$ solution, in section 4.1 of \cite{Bobev:2020fon}, and the discussion of the one-point function follows the same structure as for the $\mathcal N=2$ interface.
The $\mathcal N=1$ solution also uses a radial coordinate $X$ satisfying (\ref{eq:X-eq}), with metric in (\ref{eq:5dmetric-N2Janus}), where the warp factor and effective potential are now given by
\begin{align}
	\mathrm e^{2A}&=\frac{5^{5/3}\mathcal I}{36}\mathrm e^{2X}~,
	\nonumber\\
	V_{\rm eff}(X) &=4\mathrm e^{-2X} \left[\frac9{5^{5/3}\mathcal I} -\mathrm e^{-4X}\cosh^2 3X\right]~,
	\label{eq: warp-Veff-N1}
\end{align}
with integral of motion $0<\mathcal{I}\leq 1$.  
The turning point where the potential vanishes is $X_{\rm tp}$.
The AdS$_5$ regions now correspond to $X\rightarrow\infty$.
The 5d dilaton $\varphi$ is given by
\begin{equation}
	\cosh 2\varphi=\cosh 2F+\frac12 \mathrm{e}^{-2F}\mathcal J^2 ~,
	\label{eq: varphi-FN1}
\end{equation}
where $\mathcal J$ is a second integral of motion, and
\begin{align}
	F(X)&=F_0\pm\int_{X_{\rm tp}}^X \!\dd x\, \frac{9\mathrm e^{-x}}{5^{5/6}\sqrt{\mathcal I}\cosh 3x}
	\frac1{\sqrt{-V_{\rm eff}(x)}}~.
	\label{eq: F-intN1}
\end{align}
We can again choose the frame \eqref{eq: frame}, where the transformed 5d dilaton is $\varphi= F$. 
With $\varphi_\pm$ denoting the two asymptotic values of the 5d dilaton, the difference is
\begin{align}
	\eta\equiv \frac{\left|\varphi_+-\varphi_-\right|}{2} =\int_{X_{\rm tp}}^\infty\!\dd x\,\frac{9\mathrm e^{-x}}{5^{5/6}\sqrt{\mathcal I}\cosh 3x}\frac1{\sqrt{-V_{\rm eff}(x)}}~.
	\label{eq: etaN1}
\end{align}
The integration endpoints follow from the fact that both asymptotic AdS$_5$ regions correspond to $X\rightarrow\infty$, while the two branches meet at $X=X_{\rm tp}$. We did not solve this integral in terms of elementary functions, but it is straightforward to evaluate numerically.

The 10d uplift of the axion and dilaton simplifies, and is given by  \cite{Bobev:2020fon}
\begin{align}
	\tau=C_0+i\mathrm e^{-\Phi}=\frac{i-\sinh2\varphi\,\sin c}{\cosh2\varphi-\sinh2\varphi\,\cos c}~.
	\label{eq: upliftN1}
\end{align}
In particular, it has no dependence on $(\chi,\omega)$ or on the coordinates of the
internal space. Moreover, choosing the frame \eqref{eq: frame}\footnote{Here, we have chosen $c=\pi$. This is equivalent and yields the desired sign.}, we immediately get
\begin{align}
	\tau&=i\mathrm e^{-2\varphi}
	&
	\Phi&=2\varphi~.
	\label{eq: final-dilatonN1}
\end{align}
Thus, in the $\mathcal N=1$ solution, the 5d frame choice \eqref{eq: frame} immediately uplifts to a 10d frame with vanishing axion, and no further $SL(2,R)$ transformation is required. Importantly, \eqref{eq: final-dilatonN1} is an exact identity valid throughout the bulk, it is not obtained from an expansion around the asymptotic $\mathrm{AdS}_5$ regions.

\paragraph{One-point function} To derive the one-point function we again start with the expansion of the 5d dilaton near the asymptotic regions, which can be obtained from the integrand of \eqref{eq: F-intN1} as
\begin{align}
	\varphi(r\to\pm\infty)=\varphi_\pm \mp\frac9{2\,5^{5/6}\sqrt{\mathcal I}}\mathrm e^{-4X}+O(\mathrm e^{-6X})~.
	\label{eq: varphi-expN1}
\end{align}
Through the uplift this yields
\begin{align}
	\Phi=\Phi_\pm \mp\frac9{5^{5/6}\sqrt{\mathcal I}}\mathrm e^{-4X}+O(\mathrm e^{-6X})~,
	\label{eq: Phi-expN1}
\end{align}
From the warp factor \eqref{eq: warp-Veff-N1}
\begin{align}
	f^4=z^4\mathrm e^{-4A}=\frac{1296\, z^4 \mathrm e^{-4X}}{\mathcal I^2 5^{10/3}} \qquad \Longrightarrow\qquad \mathrm e^{-4X}=\frac{\mathcal I^2 5^{10/3}}{1296\,z^4}f^4+O(f^6)~,
	\label{eq: X-FGN1}
\end{align}
we find
\begin{align}
	\mathrm e^{\Phi}
	=\mathrm e^{\Phi_\pm}
	\left[1\mp\frac{\mathcal I^{3/2} 5^{5/2}}{144\,z^4}f^4+O(f^6)\right]~.
	\label{eq: expPhi-expN1}
\end{align}
Comparing \eqref{eq: FGexp} with \eqref{eq: expPhi-expN1}, and using $z=|x^3|$, we finally obtain
\begin{align}
	\langle {\mathcal O_{\mathcal L}}\rangle=-\frac{N^2 5^{5/2} \mathcal I^{3/2}}{288\pi^2}\frac{\sgn(x^3)}{|x^3|^4}  ~.
	\label{eq: 1ptL}
\end{align}
This result is exact in the parameter $\mathcal I$. To express it in terms of the physical coupling jump $\gamma$, one has to determine  $\eta(\mathcal I)$ from \eqref{eq: etaN1}, which in turn yields $\gamma$ through the relation
\begin{align}
	\gamma=\tanh\bigl(2\eta(\mathcal I)\bigr)~.
\end{align}

In summary, the algorithm for producing pairs of $\gamma=(\lambda_L-\lambda_R)/(\lambda_L+\lambda_R)$ and one-point functions is to fix a real value for $X_{\rm tp}$ with $\frac{\log 5}{6}<X_{\rm tp}<\infty$, and then determine $\gamma$ and the one-point function as
\begin{align}\label{eq:N1Janus-cO-sum}
	\gamma&=\tanh \left[\int_{X_{\rm tp}}^\infty\frac{3\dd x}{\cosh (3x)\sqrt{\mathrm e^{4(X_{\rm tp}-x)}\cosh^2(3x)\sech^2(3X_{\rm tp})-1}}\right]~,
	\nonumber\\[0.5em]
	\langle {\mathcal O_{\mathcal L}}\rangle&=-\frac{3N^2}{32\pi^2}\frac{e^{6X_{\rm tp}}}{\cosh^3(3X_{\rm tp})}\frac{\sgn(x^3)}{|x^3|^4}  ~.
\end{align}
With both quantities expressed in terms of one parameter, this gives an implicit relation between $\mathcal C_{\mathcal O_{\mathcal L}}$ and  $\gamma$.

\subsubsection{Non-susy Janus}

The one-point functions for non-supersymmetric Janus has been computed in \cite{Clark:2004sb}, with a typo which we shall correct for the comparison to the free energy below.
The solution is given directly in terms of 10d variables as
\begin{align}\label{eq:N0Janus-metric}
ds_{10}^2 &= L^2e^{2A(\mu)}\left(d\mu^2+ds_{\ads_4}^2\right)+L^2 ds^2_{S^5}~,
\end{align}
where, with $'$ denoting derivatives with respect to $\mu$, the warp factor and dilaton are determined in terms of a parameter $c$ by
\begin{align}\label{eq:N0Janus-dilaton}
A'(\mu)^2 &= e^{2A(\mu)}-1+\frac{c^2}{24}e^{-6A(\mu)}~,
&
\Phi'(\mu) &= c\,e^{-3A(\mu)}~.
\end{align}
Similar to the $\mathcal N=1$ interface, the one-point function has a simple expression in terms of an auxiliary parameter, which in turn sets the coupling jump in a non-trivial way. The one-point function is expressed in terms of a parameter $c$ as \cite[(3.7)]{Clark:2004sb}
\bea \label{eq: one-point-nonsusy}
\cC_{\mathcal O_{\mathcal{L}}}^R=\frac{cN^2}{8\pi^2}~.
\eea
Following \cite[(2.17)]{Clark:2004sb}, the coupling jump $\gamma$ can be expressed in terms of the deformation parameter $c$ as
\begin{align} \label{eq: delta-expansion}
\gamma&=\tanh \delta~,
	&
\delta &=\frac{c\sqrt{\pi}}{2}
\sum_{n=0}^{\infty}
\frac{\Gamma(4n+2)}
{\Gamma\!\left(3n+\frac{5}{2}\right)n!}
\left(\frac{c^2}{24}\right)^n~.
\end{align}
One may express $\gamma$ in terms of generalized hypergeometric functions as 
\begin{align}
	\gamma&=\tanh\left[\frac{2c}{3} \,
	_4F_3\left(\frac{1}{2},\frac{3}{4},1,\frac{5}{4};\frac{5}{6},\frac{7}{6},\frac{3}{2};\frac{32}{81}c^2\right)\right]~.
\end{align}
To get an explicit expression for $\cC_{\mathcal O_{\mathcal{L}}}$ in terms of $\gamma$ one can invert the relation between $c$ and $\gamma$ numerically or in a small-$\gamma$ expansion.\footnote{We note that the resulting expansion disagrees with the small-$c$ expansion in (2.18) in \cite{Clark:2004sb} at $\mathcal O(c^3)$, though we agree with (2.17) and with the expression in (2.18) before expanding.}

\subsubsection{Free energies}

The free energy can in principle be computed directly in the 5d supergravity description as finite part of an entanglement entropy, without going through the uplift to 10d.\footnote{The entanglement entropy can be described by an 8d RT surface in 10d or a 3d RT surface in 5d. The difference is an integration over the internal space in 10d, whose only effect is to convert the 10d Newton constant to the effective 5d Newton constant.} Identifying the 5d gravity parameters with field theory quantities, however, needs the uplift.

In order to compute the entanglement entropy of a spherical region centered on the defect it is convenient to choose the $AdS_4$ metric as
\begin{equation}
	ds^2_{\ads_4} =\frac{dz^2-dt^2+du^2+u^2 d\theta^2}{z^2}~.
\end{equation}
Then the RT surface for the entanglement entropy of a spherical region of radius $R$ centered on the interface is given by
\begin{align}
t&=0~, &z^2+u^2&=R^2~.
\end{align}
and with $(z,u)=(R\sech\eta,R\tanh\eta)$ the induced metric on the surface becomes
\begin{equation}
ds^2_{\gamma}=dr^2+e^{2A(r)}(d\eta^2+\sinh^2\eta d\theta^2)~.
\end{equation}
So the entanglement entropy is
\begin{equation}
S_{\rm EE}=\frac{\text{Area}(\gamma)}{4G_5}=\frac{1}{4G_5}\vol (H^2)\int dr\, e^{2A(r)}=-\frac{\pi}{2G_5}\int dr\, e^{2A(r)}~.
\end{equation}
where we used the renormalized volume of $H^2$. The defect free energy is obtained by subtracting the entanglement entropy for a theory without interface. This leads to
\begin{equation} \label{eq: def-holo-gen}
\mathcal F_{\rm def}=\frac{\pi}{2G_5}\left[\int_{-r_c'}^{+r_c'} dr e^{2A(r)}-\int_{-r_c}^{+r_c} dr \cosh^2 r \right]\,,
\end{equation}
where $\cosh^2 r$ is the warp factor of undeformed AdS and the cut-offs $r_c$, $r_c'$ are related by the requirement that the induced metric on the cut-off surface be identical,
\begin{equation}\label{eq:cutoffs}
	e^{2A(r_c')}=\cosh^2 r_c~.
\end{equation}
In terms of the radial coordinate $X$ satisfying (\ref{eq:X-eq}) the defect free energy becomes
\begin{equation} \label{eq:Fdef-X}
	\mathcal F_{\rm def}=\frac{\pi}{2G_5}\left[\int_{-X_c}^{+X_c} dX \frac{e^{2A(X)}}{\sqrt{-\vef(X)}}-\int_{-r_c}^{+r_c} dr \cosh^2 r \right]\,.
\end{equation}
We will evaluate this free energy for the $\mathcal N=2$ Janus interface, which will show that the relation between $\langle \mathcal O_{\mathcal L}\rangle$ and the derivative of the defect free energy in \eqref{eq: dF-to-L} holds. For the $\mathcal N=1,0$ interfaces we will also demonstrate the relation \eqref{eq: dF-to-L}. To this end it, it will be useful to rephrase the relation \eqref{eq: dF-to-L} as derivative with respect to the coupling jump $\gamma$, as
\begin{equation} \label{eq: one-point-delta}
\cC_{\mathcal O_{\mathcal{L}}}^R =\frac{3}{8\pi^2}(1-\gamma^2)\frac{d}{d\gamma } \mathcal F_{\rm def}
=\frac{3}{8\pi^2}\frac{d}{d\delta }\mathcal F_{\rm def}~.
\end{equation}
The first equality follows from $\lambda_{\pm}\p_{\lambda_{\pm}}\gamma = \pm \frac{1-\gamma^2}{2}$, where $\lambda_\pm = \lambda_{L/R}$, and the second equality expresses the derivative in terms of half the jump of the 10d dilaton $\delta$,
\begin{align}
	\delta &=\frac{\Phi_+ - \Phi_ -}{2}~, & \gamma &=\tanh \delta~.
\end{align}

\paragraph{$\mathcal N=2$ Janus:} For $\mathcal N=2$ Janus, with effective potential in (\ref{eq: Veff}) and warp factor in (\ref{eq: warp}), the defect free energy becomes
\begin{equation}
\frac{\mathcal F_{\rm def}}{N^2}=\frac{\mathcal{I}^{3/4}}{4}
\int_{X_c}^{X_{\mathrm{tp}}}
\frac{dX}
{X^{3/2}\sqrt{\sqrt{\mathcal{I}}-2\sqrt{X(1-X)}}}-\int _{-r_c}^{r_c} dr \cosh^2 r~,
\end{equation}
with the cut-offs related by (\ref{eq:cutoffs}). Evaluating the one-dimensional integral numerically shows that 
\begin{equation}
	\mathcal F_{\rm def}=-\frac{1}{4}N^2\log (1-\mathcal I)~.
\end{equation}
Upon using the identification $I=\gamma^2$ following from (\ref{eq:gamma-N2}), this agrees with the result for the $\mathcal N=4$ Janus interface, again as expected based on the localization discussion.

\paragraph{$\mathcal N=1$ Janus:} For $\mathcal N=1$ Janus, where $V_{\rm eff}$ was given in (\ref{eq: warp-Veff-N1}), the defect free energy becomes
\begin{equation}
	\mathcal F_{\rm def}=N^2\left[\frac{5^{5/3}}{18} \mathcal{I} \int _{X_{\rm tp}}^{X_c} \frac{e^{2X}}{\sqrt{-\vef(X)}} dX-\int_{-r_c}^{+r_c} dr \cosh^2 r \right]\,,
\end{equation}
with the cut-offs related as before by (\ref{eq:cutoffs}). Instead of evaluating the expression explicitly, we will demonstrate the relation to the one-point function indirectly.
Thanks to the uplift relation for the $\mathcal N=1$ interface, $\delta=2\eta$ with $\eta$ given in \eqref{eq: etaN1}. It will be convenient to formulate the relation (\ref{eq: one-point-delta}) in terms of $\mathcal I$, which controls the coupling jump, leading to
\bea
\frac{1}{N^2}\deri{\mathcal F_{\rm def}}{\mathcal I}=\frac{5^{5/2}\mathcal I^{3/2}}{54}\frac{d \eta}{d\mathcal I}~,
\eea
where (\ref{eq: 1ptL}) was also used.
The derivatives of $\mathcal F_{\rm def}$ and $\eta$ can be evaluated noting that the cut-off $X_c$ and turning point $X_{\rm tp}$ both depend on $\mathcal {I}$.
Defining $X_{\epsilon}=X_{\rm tp}+\epsilon$ as a regulator for the divergence of $\frac{1}{\sqrt{-\vef (X)}}$ at the turning point, we have
\begin{align}
\deri{\mathcal F_{\rm def}}{\mathcal I}&=N^2\left[\int \frac{\partial}{\partial \mathcal{I}}\left[\frac{5^{5/3}\mathcal{I}e^{2X}}{18\sqrt{-V_{\mathrm{eff}}}}\right] dX-\frac{5^{5/3}\mathcal{I}e^{2X}}{18\sqrt{-V_{\mathrm{eff}}}}\big\vert_{X=X_\epsilon}\frac{dX_{\mathrm{tp}}}{d\mathcal{I}}+\frac{5^{5/3}\mathcal{I}e^{2X}}{18\sqrt{-V_{\mathrm{eff}}}}\big\vert_{X=X_c}\frac{dX_{c}}{d\mathcal{I}}\right]
\nonumber\\
\frac{d \eta}{d \mathcal I}&=\int \frac{\partial}{\partial\mathcal{I}}\left[\frac{9e^{-X}}{5^{5/6}\sqrt{\mathcal{I}}\cosh(3X)\sqrt{-V_{\mathrm{eff}}}}\right] dX-\frac{9e^{-X}}{5^{5/6}\sqrt{\mathcal{I}}\cosh(3X)\sqrt{-V_{\mathrm{eff}}}}\big\vert_{X=X_\epsilon}\frac{dX_{\mathrm{tp}}}{d\mathcal{I}}
\end{align}
Using the relations
\begin{align}
	\frac{dX_c}{d\mathcal{I}} &=-\frac{1}{2\mathcal{I}}~,
	&
	\frac{dX_{\mathrm{tp}}}{d\mathcal{I}}&=
	-\frac{1}{2\mathcal{I}\left[3\tanh(3X_{\mathrm{tp}})-2\right]}~,
\end{align}
where $X_c (\mathcal I)$ is determined from the cut-off relation to the undeformed theory and $X_{\rm tp}$ from $\vef(X_{\mathrm{tp}})=0$, along with
\begin{equation}
\frac{\partial}{\partial\mathcal{I}}
\left[
\frac{5^{5/3}\mathcal{I}e^{2X}}
{18\sqrt{-V_{\mathrm{eff}}}}
\right]
-
\frac{5^{5/2}\mathcal{I}^{3/2}}{54}
\frac{\partial}{\partial\mathcal{I}}
\left[
\frac{9e^{-X}}
{5^{5/6}\sqrt{\mathcal{I}}\cosh(3X)\sqrt{-V_{\mathrm{eff}}}}
\right]
=
\frac{\partial}{\partial X}
\left[
\frac{5^{5/3}e^{2X}}
{36\sqrt{-V_{\mathrm{eff}}}}
\right],
\end{equation}
we arrive at 
\begin{align}
\frac{1}{N^2}
\deri{\mathcal F_{\rm def}}{\mathcal I}-\frac{16\pi^2}{3 N^2}\cC_{\mathcal O_{\mathcal {L}}}^R \frac{d \eta}{d\mathcal I}=\,&\frac{5^{5/3}e^{2X}}
{36\sqrt{-V_{\mathrm{eff}}}}\Big\vert^{X_c}_{X_{\rm tp}}
-\frac{5^{5/3}e^{2X}}{36\sqrt{-V_{\mathrm{eff}}}}\Big\vert_{X=X_c}
\nonumber\\
&-\left(\frac{5^{5/3}\mathcal{I}e^{2X_{\mathrm{tp}}}}{18}-\frac{5^{5/3}\mathcal{I}e^{2X_{\mathrm{tp}}}}{18}\,3\left[1-\tanh\left(3X_{\mathrm{tp}}\right)\right]\right)\frac{dX_{\mathrm{tp}}}{d\mathcal{I}}
\nonumber\\
&=\frac{5^{5/3}e^{2X}}
{36\sqrt{-V_{\mathrm{eff}}}}\Big\vert^{X_c}_{X_{\rm tp}}-\frac{5^{5/3}e^{2X}}{36\sqrt{-V_{\mathrm{eff}}}}\Big\vert_{X=X_c}+\frac{5^{5/3}e^{2X}}{36\sqrt{-V_{\mathrm{eff}}}}\Big\vert_{X_{\rm tp}}
\nonumber\\
&=0~.
\end{align}
This shows that the relation is indeed satisfied.

\paragraph{Non-susy Janus:} For the non-supersymmetric Janus interface we convert the solution as given in (\ref{eq:N0Janus-metric}), (\ref{eq:N0Janus-dilaton}) to the form \eqref{eq: metric-AdS5-AdS4} with radial coordinate $X$. To this end we set
\begin{align}
	dr&=e^{A}d\mu~, & X&=e^{-2A}~.
\end{align}
The radial coordinate $X$ then satisfies (\ref{eq:X-eq}) and the solution is defined by
\begin{align}  \label{eq: nonsusy-X-equations}
\dot X^{\,2}&=-V_{\rm eff}(X)~, & V_{\rm eff}(X)&=-4X^2f(X)~, & f(X)&=1-X+\frac{c^2}{24}X^4~,
\nonumber\\
\dot\Phi&=cX^2~, & \dot A^{\,2}&=f(X)~,
\end{align}
where $\cdot$ denotes derivatives with respect to $r$, and $0<c<\frac{9}{4\sqrt{2}}$ ($c=0$ corresponding to pure AdS).
The map from $r$ to $X$ has two branches.
On each branch, $0<X<X_{\rm tp}$ where the asymptotic boundary is at $X=0$.

After this conversion we can directly use the defect free energy in (\ref{eq:Fdef-X}), which becomes
\begin{equation}
	\mathcal F_{\rm def}=N^2\left[\int_{X_c}^{X_{\rm tp}}
	\frac{dX}{X^2\sqrt{f(X)}}-\int_{-r_c}^{+r_c} dr \cosh^2 r \right]\,,
\end{equation}
where the identification of the cut-offs is
\bea
X_c^{-1}=\cosh^2 r_c~.
\eea
For the half difference between the asymptotic dilaton values we find
\begin{equation}
	\delta  = \int \frac{\deri{\Phi}{r}}{\sqrt{-\vef(X)}}   dX=
	\frac{c}{2}\int_0^{X_{\rm tp}}
	\frac{X\,dX}{\sqrt{f(X)}}=\frac{c\sqrt{\pi}}{2}
	\sum_{n=0}^{\infty}
	\frac{\Gamma(4n+2)}
	{\Gamma\!\left(3n+\frac52\right)n!}
	\left(\frac{c^2}{24}\right)^n~.
\end{equation}
The relation between the one-point function and the defect free energy (\ref{eq: dF-to-L}), converted to $c$ as intermediate variable and using (\ref{eq: one-point-nonsusy}), becomes
\begin{equation} \label{eq: free-relation-nonsusy}
	\frac{1}{N^2}\deri{\mathcal F_{\rm def}}{c}
	=\frac{c}{3}\deri{\delta}{c}~.
\end{equation}
We will now show that this relation is satisfied.

Again we introduce $X_{\varepsilon}=X_{\rm tp}-\varepsilon$ to regulate the divergence of $\frac{1}{\sqrt{f(X_{\rm tp})}}$ at the turning point.
Taking a derivative with respect to $c$ then yields
\bea  \label{eq: nonsusy-derivatives}
 \deri{\mathcal F_{\rm def}}{c}
 &=N^2\left[\int \partial_c\left[
 \frac{1}{X^2\sqrt{f(X)}}
 \right]dX
 +\left.
 \frac{1}{X^2\sqrt{f(X)}}
 \right|_{X=X_\varepsilon} \deri{X_{\rm tp}}{c}\right]~,
 \\
 \deri{\delta}{c}
 &=\int \partial_c\left[
 \frac{cX}{2\sqrt{f(X)}}
 \right]dX
 +\left.
 \frac{cX}{2\sqrt{f(X)}}
 \right|_{X=X_\varepsilon}\deri{X_{\rm tp}}{c}~.
\eea
Using the relations
\begin{align}
 \partial_c\left(\frac{1}{X^2\sqrt f}\right)
 -\frac{c}{3}\,
 \partial_c\left(\frac{cX}{2\sqrt f}\right)
 &=-\frac{c}{12}\,
 \partial_X\left(\frac{X^2}{\sqrt f}\right)~,
 &
  2\deri{X_{\rm tp}}{c}
 &=\frac{cX_{\rm tp}^4}
 {6-c^2X_{\rm tp}^3}~,
\end{align}
where the second equation is obtained from $f(X_{\rm tp})=0$, we obtain
\bea
\frac{1}{N^2}
\deri{\mathcal F_{\rm def}}{c}-\frac{c}{3}\deri{\delta}{c}
&\left.=-\frac{c}{12}\frac{X^2}{\sqrt{f(X)}}\right|^{X_{\varepsilon}}_{X_c}
+\deri{X_{\rm tp}}{c}\frac{1}{\sqrt{f(X)}}(\frac{1}{X^2}-\left.\frac{c^2}{6}X)\right|_{X_{\varepsilon}}\\
&\left.=\frac{c}{12}\frac{X^2}{\sqrt{f(X)}}\right|_{X_c}=0~.
\eea
This shows that (\ref{eq: free-relation-nonsusy}) holds.
From \eqref{eq: delta-expansion} and \eqref{eq: free-relation-nonsusy}, we can also obtain $\mathcal F_{\rm def}$ as a series expansion in $c$
\bea
\mathcal F_{\rm def}=N^2 \frac{c^2\sqrt{\pi}}{24}
\sum_{n=0}^{\infty}
\frac{\Gamma(4n+3)}
{\Gamma\left(3n+\frac{5}{2}\right)(n+1)!}
\left(\frac{c^2}{24}\right)^n~.
\eea
In terms of generalized hypergeometric functions this can be expressed as
\begin{equation}
	\mathcal F_{\rm def}=\frac{c^2}{9}N^2  \,
	_4F_3\left(\frac{3}{4},1,1,\frac{5}{4};\frac{5}{6},\frac{7}{6},2;\frac{32}{81}c^2\right)~.
\end{equation}

We can draw a humble lesson from the fact that the relation (\ref{eq: dF-to-L}) between the defect free energy and the Lagrangian one-point function holds for the non-supersymmetric Janus solution: A Lagrangian for the dual theory was proposed in \cite{Clark:2004sb}, in the form of the ambient Lagrangian corresponding to $\mathcal O_{\mathcal L}$, i.e.\ with $\phi\square\phi$ scalar kinetic term, and no additional interface terms. With no supersymmetry one could add additional interface terms, and the conjectural nature of this proposal was emphasized in \cite{Clark:2004sb}. In view of the discussion in section \ref{sec:integrated-L}, any additional interface terms could potentially show up in the relation between the defect free energy and $\langle \mathcal O_{\mathcal L}\rangle$. The fact that this is not the case and the relation holds unmodified can therefore be seen as consistency check for the proposal in \cite{Clark:2004sb}.

\section*{Acknowledgments}	
We thank Davide Bason, Marius Gerbershagen and especially Andreas Karch for valuable discussions and thank all authors of \cite{Karch:2026ymg} for correspondence.
DH is supported by FWO-Vlaanderen projects G012222N and G0A2226N, and by the VUB Research Council through the Strategic Research Program High-Energy Physics.

\phantomsection
\addcontentsline{toc}{section}{References}
\bibliographystyle{JHEP}
\bibliography{BCFT.bib}

\providecommand{\href}[2]{#2}\begingroup\raggedright\begin{thebibliography}{10}

\bibitem{Gaiotto:2008sd}
D.~Gaiotto and E.~Witten, {\it {Janus Configurations, Chern-Simons Couplings, And The theta-Angle in N=4 Super Yang-Mills Theory}},  {\em JHEP} {\bf 06} (2010) 097, [\href{https://arxiv.org/abs/0804.2907}{{\tt arXiv:0804.2907}}].

\bibitem{Gaiotto:2008sa}
D.~Gaiotto and E.~Witten, {\it {Supersymmetric Boundary Conditions in N=4 Super Yang-Mills Theory}},  {\em J. Statist. Phys.} {\bf 135} (2009) 789--855, [\href{https://arxiv.org/abs/0804.2902}{{\tt arXiv:0804.2902}}].

\bibitem{Gaiotto:2008ak}
D.~Gaiotto and E.~Witten, {\it {S-Duality of Boundary Conditions In N=4 Super Yang-Mills Theory}},  {\em Adv. Theor. Math. Phys.} {\bf 13} (2009), no.~3 721--896, [\href{https://arxiv.org/abs/0807.3720}{{\tt arXiv:0807.3720}}].

\bibitem{Nagasaki:2011ue}
K.~Nagasaki, H.~Tanida, and S.~Yamaguchi, {\it {Holographic Interface-Particle Potential}},  {\em JHEP} {\bf 01} (2012) 139, [\href{https://arxiv.org/abs/1109.1927}{{\tt arXiv:1109.1927}}].

\bibitem{Nagasaki:2012re}
K.~Nagasaki and S.~Yamaguchi, {\it {Expectation values of chiral primary operators in holographic interface CFT}},  {\em Phys. Rev. D} {\bf 86} (2012) 086004, [\href{https://arxiv.org/abs/1205.1674}{{\tt arXiv:1205.1674}}].

\bibitem{deLeeuw:2015hxa}
M.~de~Leeuw, C.~Kristjansen, and K.~Zarembo, {\it {One-point Functions in Defect CFT and Integrability}},  {\em JHEP} {\bf 08} (2015) 098, [\href{https://arxiv.org/abs/1506.06958}{{\tt arXiv:1506.06958}}].

\bibitem{Buhl-Mortensen:2015gfd}
I.~Buhl-Mortensen, M.~de~Leeuw, C.~Kristjansen, and K.~Zarembo, {\it {One-point Functions in AdS/dCFT from Matrix Product States}},  {\em JHEP} {\bf 02} (2016) 052, [\href{https://arxiv.org/abs/1512.02532}{{\tt arXiv:1512.02532}}].

\bibitem{Buhl-Mortensen:2016pxs}
I.~Buhl-Mortensen, M.~de~Leeuw, A.~C. Ipsen, C.~Kristjansen, and M.~Wilhelm, {\it {One-loop one-point functions in gauge-gravity dualities with defects}},  {\em Phys. Rev. Lett.} {\bf 117} (2016), no.~23 231603, [\href{https://arxiv.org/abs/1606.01886}{{\tt arXiv:1606.01886}}].

\bibitem{deLeeuw:2017cop}
M.~de~Leeuw, A.~C. Ipsen, C.~Kristjansen, and M.~Wilhelm, {\it {Introduction to integrability and one-point functions in $\mathcal N=$ 4 supersymmetric Yang\textendash{}Mills theory and its defect cousin}},  \href{https://arxiv.org/abs/1708.02525}{{\tt arXiv:1708.02525}}.

\bibitem{deLeeuw:2019usb}
M.~de~Leeuw, {\it {One-point functions in AdS/dCFT}},  {\em J. Phys. A} {\bf 53} (2020), no.~28 283001, [\href{https://arxiv.org/abs/1908.03444}{{\tt arXiv:1908.03444}}].

\bibitem{DeLeeuw:2018cal}
M.~De~Leeuw, C.~Kristjansen, and G.~Linardopoulos, {\it {Scalar one-point functions and matrix product states of AdS/dCFT}},  {\em Phys. Lett. B} {\bf 781} (2018) 238--243, [\href{https://arxiv.org/abs/1802.01598}{{\tt arXiv:1802.01598}}].

\bibitem{Linardopoulos:2025ypq}
G.~Linardopoulos, {\it {String theory methods for defect CFTs}},  \href{https://arxiv.org/abs/2501.11985}{{\tt arXiv:2501.11985}}.

\bibitem{Robinson:2017sup}
B.~Robinson and C.~F. Uhlemann, {\it {Supersymmetric D3/D5 for massive defects on curved space}},  {\em JHEP} {\bf 12} (2017) 143, [\href{https://arxiv.org/abs/1709.08650}{{\tt arXiv:1709.08650}}].

\bibitem{Wang:2020seq}
Y.~Wang, {\it {Taming defects in $ \mathcal{N} $ = 4 super-Yang-Mills}},  {\em JHEP} {\bf 08} (2020), no.~08 021, [\href{https://arxiv.org/abs/2003.11016}{{\tt arXiv:2003.11016}}].

\bibitem{Komatsu:2020sup}
S.~Komatsu and Y.~Wang, {\it {Non-perturbative defect one-point functions in planar $\mathcal{N}=4$ super-Yang-Mills}},  {\em Nucl. Phys. B} {\bf 958} (2020) 115120, [\href{https://arxiv.org/abs/2004.09514}{{\tt arXiv:2004.09514}}].

\bibitem{He:2025due}
D.~He and C.~F. Uhlemann, {\it {One-point functions for doubly-holographic BCFTs and backreacting defects}},  {\em JHEP} {\bf 25} (2025), no.~5 227, [\href{https://arxiv.org/abs/2501.07630}{{\tt arXiv:2501.07630}}].

\bibitem{Clark:2004sb}
A.~B. Clark, D.~Z. Freedman, A.~Karch, and M.~Schnabl, {\it {Dual of the Janus solution: An interface conformal field theory}},  {\em Phys. Rev. D} {\bf 71} (2005) 066003, [\href{https://arxiv.org/abs/hep-th/0407073}{{\tt hep-th/0407073}}].

\bibitem{Karch:2026ymg}
A.~Karch, A.~Sanyal, R.~C. Spieler, and M.~Wang, {\it {One-point functions in 2D and 4D SUSY Janus}},  \href{https://arxiv.org/abs/2604.03185}{{\tt arXiv:2604.03185}}.

\bibitem{Chicherin:2016fbj}
D.~Chicherin and E.~Sokatchev, {\it {$ \mathcal{N} $ = 4 super-Yang-Mills in LHC superspace part II: non-chiral correlation functions of the stress-tensor multiplet}},  {\em JHEP} {\bf 03} (2017) 048, [\href{https://arxiv.org/abs/1601.06804}{{\tt arXiv:1601.06804}}].

\bibitem{Bason:2023bin}
D.~Bason, L.~Di~Pietro, R.~Valandro, and J.~van Muiden, {\it {BCFT One-point Functions of Coulomb Branch Operators}},  \href{https://arxiv.org/abs/2311.17888}{{\tt arXiv:2311.17888}}.

\bibitem{Bobev:2020fon}
N.~Bobev, F.~F. Gautason, K.~Pilch, M.~Suh, and J.~van Muiden, {\it {Holographic interfaces in $ \mathcal{N} $ = 4 SYM: Janus and J-folds}},  {\em JHEP} {\bf 05} (2020) 134, [\href{https://arxiv.org/abs/2003.09154}{{\tt arXiv:2003.09154}}].

\bibitem{DHoker:2006qeo}
E.~D'Hoker, J.~Estes, and M.~Gutperle, {\it {Interface Yang-Mills, supersymmetry, and Janus}},  {\em Nucl. Phys. B} {\bf 753} (2006) 16--41, [\href{https://arxiv.org/abs/hep-th/0603013}{{\tt hep-th/0603013}}].

\bibitem{McAvity:1995zd}
D.~M. McAvity and H.~Osborn, {\it {Conformal field theories near a boundary in general dimensions}},  {\em Nucl. Phys. B} {\bf 455} (1995) 522--576, [\href{https://arxiv.org/abs/cond-mat/9505127}{{\tt cond-mat/9505127}}].

\bibitem{Billo:2016cpy}
M.~Bill{\`o}, V.~Gon{\c{c}}alves, E.~Lauria, and M.~Meineri, {\it {Defects in conformal field theory}},  {\em JHEP} {\bf 04} (2016) 091, [\href{https://arxiv.org/abs/1601.02883}{{\tt arXiv:1601.02883}}].

\bibitem{Liendo:2012hy}
P.~Liendo, L.~Rastelli, and B.~C. van Rees, {\it {The Bootstrap Program for Boundary CFT$_d$}},  {\em JHEP} {\bf 07} (2013) 113, [\href{https://arxiv.org/abs/1210.4258}{{\tt arXiv:1210.4258}}].

\bibitem{Liendo:2016ymz}
P.~Liendo and C.~Meneghelli, {\it {Bootstrap equations for $ \mathcal{N} $ = 4 SYM with defects}},  {\em JHEP} {\bf 01} (2017) 122, [\href{https://arxiv.org/abs/1608.05126}{{\tt arXiv:1608.05126}}].

\bibitem{Karch:2000ct}
A.~Karch and L.~Randall, {\it {Locally localized gravity}},  {\em JHEP} {\bf 05} (2001) 008, [\href{https://arxiv.org/abs/hep-th/0011156}{{\tt hep-th/0011156}}].

\bibitem{Karch:2000gx}
A.~Karch and L.~Randall, {\it {Open and closed string interpretation of SUSY CFT's on branes with boundaries}},  {\em JHEP} {\bf 06} (2001) 063, [\href{https://arxiv.org/abs/hep-th/0105132}{{\tt hep-th/0105132}}].

\bibitem{Uhlemann:2021nhu}
C.~F. Uhlemann, {\it {Islands and Page curves in 4d from Type IIB}},  {\em JHEP} {\bf 08} (2021) 104, [\href{https://arxiv.org/abs/2105.00008}{{\tt arXiv:2105.00008}}].

\bibitem{Karch:2022rvr}
A.~Karch, H.~Sun, and C.~F. Uhlemann, {\it {Double holography in string theory}},  {\em JHEP} {\bf 10} (2022) 012, [\href{https://arxiv.org/abs/2206.11292}{{\tt arXiv:2206.11292}}].

\bibitem{Witten:1988ze}
E.~Witten, {\it {Topological Quantum Field Theory}},  {\em Commun. Math. Phys.} {\bf 117} (1988) 353.

\bibitem{Nekrasov:2002qd}
N.~A. Nekrasov, {\it {Seiberg-Witten prepotential from instanton counting}},  {\em Adv. Theor. Math. Phys.} {\bf 7} (2003), no.~5 831--864, [\href{https://arxiv.org/abs/hep-th/0206161}{{\tt hep-th/0206161}}].

\bibitem{Pestun:2007rz}
V.~Pestun, {\it {Localization of gauge theory on a four-sphere and supersymmetric Wilson loops}},  {\em Commun. Math. Phys.} {\bf 313} (2012) 71--129, [\href{https://arxiv.org/abs/0712.2824}{{\tt arXiv:0712.2824}}].

\bibitem{Dedushenko:2018tgx}
M.~Dedushenko, {\it {Gluing II: boundary localization and gluing formulas}},  {\em Lett. Math. Phys.} {\bf 111} (2021), no.~1 18, [\href{https://arxiv.org/abs/1807.04278}{{\tt arXiv:1807.04278}}].

\bibitem{DHoker:2007zhm}
E.~D'Hoker, J.~Estes, and M.~Gutperle, {\it {Exact half-BPS Type IIB interface solutions. I. Local solution and supersymmetric Janus}},  {\em JHEP} {\bf 06} (2007) 021, [\href{https://arxiv.org/abs/0705.0022}{{\tt arXiv:0705.0022}}].

\bibitem{DHoker:2007hhe}
E.~D'Hoker, J.~Estes, and M.~Gutperle, {\it {Exact half-BPS Type IIB interface solutions. II. Flux solutions and multi-Janus}},  {\em JHEP} {\bf 06} (2007) 022, [\href{https://arxiv.org/abs/0705.0024}{{\tt arXiv:0705.0024}}].

\bibitem{Aharony:2011yc}
O.~Aharony, L.~Berdichevsky, M.~Berkooz, and I.~Shamir, {\it {Near-horizon solutions for D3-branes ending on 5-branes}},  {\em Phys. Rev. D} {\bf 84} (2011) 126003, [\href{https://arxiv.org/abs/1106.1870}{{\tt arXiv:1106.1870}}].

\bibitem{Assel:2011xz}
B.~Assel, C.~Bachas, J.~Estes, and J.~Gomis, {\it {Holographic Duals of D=3 N=4 Superconformal Field Theories}},  {\em JHEP} {\bf 08} (2011) 087, [\href{https://arxiv.org/abs/1106.4253}{{\tt arXiv:1106.4253}}].

\bibitem{Goto:2020per}
K.~Goto, L.~Nagano, T.~Nishioka, and T.~Okuda, {\it {Janus interface entropy and Calabi's diastasis in four-dimensional $\mathcal{N}=2$ superconformal field theories}},  {\em JHEP} {\bf 08} (2020), no.~08 048, [\href{https://arxiv.org/abs/2005.10833}{{\tt arXiv:2005.10833}}].

\bibitem{Raamsdonk:2020tin}
M.~V. Raamsdonk and C.~Waddell, {\it {Holographic and localization calculations of boundary F for $ \mathcal{N} $ = 4 SUSY Yang-Mills theory}},  {\em JHEP} {\bf 02} (2021) 222, [\href{https://arxiv.org/abs/2010.14520}{{\tt arXiv:2010.14520}}].

\bibitem{He:2024djr}
D.~He and C.~F. Uhlemann, {\it {Solving $ \mathcal{N} $ = 4 SYM BCFT matrix models at large N}},  {\em JHEP} {\bf 12} (2024) 164, [\href{https://arxiv.org/abs/2409.13016}{{\tt arXiv:2409.13016}}].

\bibitem{Assel:2012cp}
B.~Assel, J.~Estes, and M.~Yamazaki, {\it {Large N Free Energy of 3d N=4 SCFTs and $AdS_4/CFT_3$}},  {\em JHEP} {\bf 09} (2012) 074, [\href{https://arxiv.org/abs/1206.2920}{{\tt arXiv:1206.2920}}].

\bibitem{Bachas:2017wva}
C.~Bachas, M.~Bianchi, and A.~Hanany, {\it {$ \mathcal{N}=2 $ moduli of AdS$_{4}$ vacua: a fine-print study}},  {\em JHEP} {\bf 08} (2018) 100, [\href{https://arxiv.org/abs/1711.06722}{{\tt arXiv:1711.06722}}]. [Erratum: JHEP 10, 032 (2018)].

\bibitem{Coccia:2020wtk}
L.~Coccia and C.~F. Uhlemann, {\it {On the planar limit of 3d $ {\mathrm{T}}_{\rho}^{\sigma}\left[\mathrm{SU}\left(\mathrm{N}\right)\right] $}},  {\em JHEP} {\bf 06} (2021) 038, [\href{https://arxiv.org/abs/2011.10050}{{\tt arXiv:2011.10050}}].

\bibitem{Chaney:2024bgx}
A.~Chaney and C.~F. Uhlemann, {\it {BMN-like sectors in 4d $\mathcal N=4$ SYM with boundaries and interfaces}},  \href{https://arxiv.org/abs/2408.12651}{{\tt arXiv:2408.12651}}.

\bibitem{Bak:2003jk}
D.~Bak, M.~Gutperle, and S.~Hirano, {\it {A Dilatonic deformation of AdS(5) and its field theory dual}},  {\em JHEP} {\bf 05} (2003) 072, [\href{https://arxiv.org/abs/hep-th/0304129}{{\tt hep-th/0304129}}].

\bibitem{Dolan:2002zh}
F.~A. Dolan and H.~Osborn, {\it {On short and semi-short representations for four-dimensional superconformal symmetry}},  {\em Annals Phys.} {\bf 307} (2003) 41--89, [\href{https://arxiv.org/abs/hep-th/0209056}{{\tt hep-th/0209056}}].

\bibitem{Basu:2004nt}
A.~Basu, M.~B. Green, and S.~Sethi, {\it {Some systematics of the coupling constant dependence of N=4 Yang-Mills}},  {\em JHEP} {\bf 09} (2004) 045, [\href{https://arxiv.org/abs/hep-th/0406231}{{\tt hep-th/0406231}}].

\bibitem{Gerchkovitz:2016gxx}
E.~Gerchkovitz, J.~Gomis, N.~Ishtiaque, A.~Karasik, Z.~Komargodski, and S.~S. Pufu, {\it {Correlation Functions of Coulomb Branch Operators}},  {\em JHEP} {\bf 01} (2017) 103, [\href{https://arxiv.org/abs/1602.05971}{{\tt arXiv:1602.05971}}].

\bibitem{Herzog:2019bom}
C.~P. Herzog and I.~Shamir, {\it {On Marginal Operators in Boundary Conformal Field Theory}},  {\em JHEP} {\bf 10} (2019) 088, [\href{https://arxiv.org/abs/1906.11281}{{\tt arXiv:1906.11281}}].

\bibitem{Gerchkovitz:2014gta}
E.~Gerchkovitz, J.~Gomis, and Z.~Komargodski, {\it {Sphere Partition Functions and the Zamolodchikov Metric}},  {\em JHEP} {\bf 11} (2014) 001, [\href{https://arxiv.org/abs/1405.7271}{{\tt arXiv:1405.7271}}].

\bibitem{Brink:1976bc}
L.~Brink, J.~H. Schwarz, and J.~Scherk, {\it {Supersymmetric Yang-Mills Theories}},  {\em Nucl. Phys. B} {\bf 121} (1977) 77--92.

\bibitem{Hanany:1996ie}
A.~Hanany and E.~Witten, {\it {Type IIB superstrings, BPS monopoles, and three-dimensional gauge dynamics}},  {\em Nucl. Phys. B} {\bf 492} (1997) 152--190, [\href{https://arxiv.org/abs/hep-th/9611230}{{\tt hep-th/9611230}}].

\bibitem{Giveon:1998sr}
A.~Giveon and D.~Kutasov, {\it {Brane Dynamics and Gauge Theory}},  {\em Rev. Mod. Phys.} {\bf 71} (1999) 983--1084, [\href{https://arxiv.org/abs/hep-th/9802067}{{\tt hep-th/9802067}}].

\bibitem{Clark:2005te}
A.~Clark and A.~Karch, {\it {Super Janus}},  {\em JHEP} {\bf 10} (2005) 094, [\href{https://arxiv.org/abs/hep-th/0506265}{{\tt hep-th/0506265}}].

\bibitem{DHoker:2006vfr}
E.~D'Hoker, J.~Estes, and M.~Gutperle, {\it {Ten-dimensional supersymmetric Janus solutions}},  {\em Nucl. Phys. B} {\bf 757} (2006) 79--116, [\href{https://arxiv.org/abs/hep-th/0603012}{{\tt hep-th/0603012}}].

\bibitem{DeWolfe:2001pq}
O.~DeWolfe, D.~Z. Freedman, and H.~Ooguri, {\it {Holography and defect conformal field theories}},  {\em Phys. Rev. D} {\bf 66} (2002) 025009, [\href{https://arxiv.org/abs/hep-th/0111135}{{\tt hep-th/0111135}}].

\bibitem{Aharony:2003qf}
O.~Aharony, O.~DeWolfe, D.~Z. Freedman, and A.~Karch, {\it {Defect conformal field theory and locally localized gravity}},  {\em JHEP} {\bf 07} (2003) 030, [\href{https://arxiv.org/abs/hep-th/0303249}{{\tt hep-th/0303249}}].

\bibitem{Beccaria:2022bjo}
M.~Beccaria and A.~Cabo-Bizet, {\it {1/N expansion of the D3-D5 defect CFT at strong coupling}},  {\em JHEP} {\bf 02} (2023) 208, [\href{https://arxiv.org/abs/2212.12415}{{\tt arXiv:2212.12415}}].

\bibitem{Alvarez-Gaume:1990asn}
L.~Alvarez-Gaume, C.~Gomez, and J.~Lacki, {\it {Integrability in random matrix models}},  {\em Phys. Lett. B} {\bf 253} (1991) 56--62.

\bibitem{Demulder:2022aij}
S.~Demulder, A.~Gnecchi, I.~Lavdas, and D.~Lust, {\it {Islands and light gravitons in type IIB string theory}},  {\em JHEP} {\bf 02} (2023) 016, [\href{https://arxiv.org/abs/2204.03669}{{\tt arXiv:2204.03669}}].

\bibitem{Billiato:2025jkr}
F.~Billiato and A.~Gnecchi, {\it {Entanglement surfaces for rotating cylindrical black holes}},  \href{https://arxiv.org/abs/2512.04193}{{\tt arXiv:2512.04193}}.

\bibitem{Coccia:2021lpp}
L.~Coccia and C.~F. Uhlemann, {\it {Mapping out the internal space in AdS/BCFT with Wilson loops}},  {\em JHEP} {\bf 03} (2022) 127, [\href{https://arxiv.org/abs/2112.14648}{{\tt arXiv:2112.14648}}].

\bibitem{Balasubramanian:1998de}
V.~Balasubramanian, P.~Kraus, A.~E. Lawrence, and S.~P. Trivedi, {\it {Holographic probes of anti-de Sitter space-times}},  {\em Phys. Rev. D} {\bf 59} (1999) 104021, [\href{https://arxiv.org/abs/hep-th/9808017}{{\tt hep-th/9808017}}].

\bibitem{Bianchi:2001kw}
M.~Bianchi, D.~Z. Freedman, and K.~Skenderis, {\it {Holographic renormalization}},  {\em Nucl. Phys. B} {\bf 631} (2002) 159--194, [\href{https://arxiv.org/abs/hep-th/0112119}{{\tt hep-th/0112119}}].

\bibitem{Skenderis:2002wp}
K.~Skenderis, {\it {Lecture notes on holographic renormalization}},  {\em Class. Quant. Grav.} {\bf 19} (2002) 5849--5876, [\href{https://arxiv.org/abs/hep-th/0209067}{{\tt hep-th/0209067}}].

\end{thebibliography}\endgroup
	
\end{document}